\documentclass[prd,letter,twocolumn,tightenlines]{revtex4}
\usepackage[hypertex]{hyperref}
\usepackage{natbib,ifthen}

\usepackage{amssymb}
\usepackage{amsmath}
\usepackage{graphicx}
\usepackage{color}
\def\beq{\begin{equation}}
\def\eeq{\end{equation}}
\def\bea{\begin{eqnarray}}
\def\eea{\end{eqnarray}}

\def\lsim{\mathrel{\mathstrut\smash{\ooalign{\raise2.5pt\hbox{$<$}\cr\lower2.5pt\hbox{$\sim$}}}}}
\def\gsim{\mathrel{\mathstrut\smash{\ooalign{\raise2.5pt\hbox{$>$}\cr\lower2.5pt\hbox{$\sim$}}}}}
\def\bwt{\begin{widetext}}
\def\ewt{\end{widetext}}

\begin{document}

\bibliographystyle{prsty}
\title{Observational Evidence for Cosmological-Scale Extra Dimensions}
\author{Niayesh Afshordi}
\author{Ghazal Geshnizjani}
\author{Justin Khoury}
 \affiliation{Perimeter Institute for Theoretical Physics, Waterloo, Ontario
N2L 2Y5, Canada}

\begin{abstract}
We present a case that current observations may already indicate new gravitational physics
on cosmological scales. The excess of power seen in the Lyman-$\alpha$ forest and small-scale CMB experiments, the anomalously large
bulk flows seen both in peculiar velocity surveys and in kinetic SZ, and the higher ISW cross-correlation all indicate that
structure may be more evolved than expected from $\Lambda$CDM. We argue that these observations find a natural explanation
in models with infinite-volume (or, at least, cosmological-size) extra dimensions, where the graviton is a resonance with a tiny width.
The longitudinal mode of the graviton mediates an extra scalar force which speeds up structure formation at late times,
thereby accounting for the above anomalies. The required graviton Compton wavelength is relatively small compared to the present Hubble radius,
of order 300-600 Mpc. Moreover, with certain assumptions about the behavior of the longitudinal mode on super-Hubble scales, our modified gravity framework can
also alleviate the tension with the low quadrupole and the peculiar vanishing of the CMB correlation function on large angular scales, seen both in COBE and WMAP. This relies on a novel mechanism that cancels a late-time ISW contribution against the primordial Sachs-Wolfe amplitude.
\end{abstract}
\maketitle

\section{Introduction}
\label{intro}

In the ten years since the discovery of cosmic acceleration by the supernovae teams~\cite{sn1a}, the $\Lambda$-cold dark matter ($\Lambda$CDM) model
has emerged as the standard paradigm for cosmology. With a single parameter ($\Lambda$), the model predicts expansion and growth histories
consistent with observations. As near-future experiments will soon probe the large scale structure with unprecedented accuracy, it is worth asking at this
juncture whether ten years from now the $\Lambda$CDM model will be hailed as the ultimate cosmological theory, or whether it will have by then joined its
Einstein de Sitter predecessor in the ranks of defunct cosmologies. A host of recent observations may have already revealed cracks in $\Lambda$CDM's armor:

\begin{enumerate}

\item The cross-correlation between galaxy surveys and the cosmic microwave background (CMB), resulting from the Integrated Sachs-Wolfe (ISW) effect,
has recently been measured at the $4\sigma$ level using the Wilkinson Microwave Anisotropy Probe (WMAP)
data and a combination of large-scale structure surveys~\cite{seljak,Giannantonio:2008zi}.
This provides non-trivial and independent evidence for cosmic acceleration. Remarkably, the combined amplitude
is larger than the $\Lambda$CDM prediction by $\gsim 2\sigma$'s: $2.23\pm 0.60$~\cite{seljak}, with 1 being the $\Lambda$CDM expectation.

\item Independent recent analyses suggest anomalously large bulk flows on large scales. Using a compilation of peculiar velocity surveys,
Watkins {\it et al.}~\cite{hudson} obtained a bulk flow of $407\pm 81$~km~s$^{-1}$ on 50$h^{-1}$~Mpc scales, in contrast with the $\Lambda$CDM expectation of
$\sim 190$~km~s$^{-1}$ for the r.m.s. value of the bulk flow, which are inconsistent at the $2\sigma$ level. Meanwhile, using the kinetic Sunyaev-Zeldovich effect to estimate peculiar velocities of galaxy clusters, Kashlinsky {\it et al.}~\cite{kash} found a coherent bulk motion of 600-1000~km~s$^{-1}$ out to $\gsim 300h^{-1}$~Mpc.

\item High resolution observations of the CMB at $\ell \sim 3000$ by the Cosmic Background Imager (CBI) indicate an excess power compared to the theoretical primary CMB power spectrum \cite{Readhead:2004gy}. This excess could be explained by the thermal Sunyaev-Zel'dovich effect from galaxy clusters at $z\sim 1$. However, the amplitude of matter power spectrum, necessary to explain CBI excess is about $35\%$ (at $\sim 2\sigma$ level) larger than the current best-fit concordance $\Lambda$CDM model \cite{Reichardt:2008ay}.

\item The Lyman-$\alpha$ forest in the spectra of high redshift quasars provides us with the highest resolution precision measurement of the cosmological matter power spectrum. However, Lyman-$\alpha$ forest constraints on the linear matter power spectrum, based on the SDSS quasar spectra at $z \sim 3$ \cite{McDonald:2004eu,McDonald:2004xn} is also $\sim 35\%$ (at $\sim 2.5\sigma$ level) larger than the $\Lambda$CDM prediction (Fig. 1 in \cite{Seljak:2006bg}).

\item As first observed by the Cosmic Background Explorer (COBE)~\cite{cobe} and later confirmed by WMAP~\cite{wmap1},
the CMB temperature anisotropy shows a lack of correlation on large ($\gsim 60\,^{\circ}$)
angular separations. While related to the low quadrupole, the real-space anomaly is more robust statistically
--- less than 0.03\% of random realizations exhibit smaller power on those scales~\cite{wmap1,huterer,others}.
The statistical significance of this anomaly has only grown with subsequent WMAP data releases~\cite{huterer2}.

\end{enumerate}

Of course, these anomalies may eventually go away, as systematic effects are better understood. After all, these represent for the most part $\sim 2\sigma$ discrepancies.
But it is striking that the first four observations on this list of misfits all point towards the same physics: that structure is more evolved on large scales than
predicted by $\Lambda$CDM. This suggests that gravity may be stronger at late times and on large scales.

In this paper we show that this is precisely what happens in infrared-modified gravity theories, inspired by brane-world constructions with
infinite-volume extra dimensions.  Because the extra dimensions are infinite in extent (at least cosmologically large), the 4d graviton is no longer massless but is
instead broadened into a resonance --- a continuum of massive states --- with a tiny width $r_c^{-1}$. At first sight this may seem at odds with our earlier conclusion, since
a massive/resonance graviton implies weaker gravity. While this is certainly true on the largest distances, on intermediate (albeit cosmologically-relevant) scales the longitudinal (or helicity-0) mode mediates an extra scalar force which enhances gravitational attraction by order unity.

This extra force is of course at the origin of the well-known van Dam-Veltman-Zakharov (vDVZ) discontinuity~\cite{vDVZ}. As conjectured by Vainshtein~\cite{vainshtein}, however,
non-linear interactions can suppress the effects of the longitudinal mode near astrophysical sources. This screening is also at play cosmologically,
suppressing the extra force at early times and on small scales. Thus gravity becomes stronger only at late times and on sufficiently large scales.
This suppression at high density or curvature is qualitatively similar to the chameleon mechanism \cite{cham1,cham2,cham3}, except that the Vainshtein effect arises from derivative interactions as opposed to a scalar potential.

The theories of interest are higher-dimensional generalizations of the Dvali-Gabadadze-Porrati (DGP) model~\cite{DGP},
in which our visible universe is confined to a 3-brane. The confrontation of its cosmological predictions with observations
has been the subject of much literature since its advent~\cite{lue,mustafa,koyama,hu,hu2,wiley0,amin,song,wiley}.
Instead of having one extra dimension, here the bulk consists of 6 or more space-time dimensions.

Extending the DGP scenario to higher dimensions has proven difficult historically.
Early studies demonstrated that naive extensions are plagued by ghost-like instabilities,
even around flat space~\cite{sergei,gigashif}.
Recently, however, it was shown that such instabilities are absent if our 3-brane lies within a succession of higher-dimensional branes, each with their own induced gravity term, and embedded in one another in a flat bulk space-time~\cite{oriol,us}. We refer to this framework as Cascading Gravity~\cite{claudiareview}. In the simplest codimension-2 case, for instance, our 3-brane is embedded in a 4-brane within a 6-dimensional bulk.
A similar cascading behavior of the gravitational force law was also obtained in a different codimension-2 framework~\cite{nemanjacharting}.

Due to the higher-dimensional nature of these constructions, extracting cosmological predictions presents a daunting technical challenge.
Even deriving the modified Friedmann equation is highly non-trivial. Here we circumvent these difficulties by developing a phenomenological description for the background cosmology, while using the parametrized
post-Friedmann (PPF)~\cite{ppf} framework to encode modifications to Einstein gravity into the evolution of perturbations. This approach has been shown to accurately reproduce
the cosmological behavior of the standard DGP model. Inspired by these results, we can infer the parametric dependence of the required
PPF input functions and parameters for Cascading Gravity models. This allows us to calculate various cosmological observables,
all within a 3+1-dimensional framework.

One of our key observations is that in 6 or more dimensions the background evolution becomes nearly indistinguishable from
$\Lambda$CDM. This allows us to consider relatively small graviton Compton wavelengths,
{\it e.g.} $r_c\sim  300-600$~Mpc, without violating constraints on the expansion history.
Compared to the usual assumption $r_c\sim H_0^{-1}$ in DGP, this expands considerably
the duration and range of scales over which enhanced gravity is effective.

As mentioned earlier, the culprit for most of the interesting phenomenology in these models is the longitudinal or helicity-0 mode of the massive graviton.
This degree of freedom is ultimately responsible for addressing all of the aforementioned anomalies:

\begin{itemize}

\item The longitudinal degree of freedom modifies the time-evolution of the lensing potential, which results in a stronger
ISW cross-correlation compared to $\Lambda$CDM.

\item In the non-relativistic limit, the helicity-0 mode can be thought of as mediating an extra scalar force, which enhances gravitational attraction on intermediate scales.
This speed-up in structure formation at late times results in larger bulk flows, more clusters at $z\sim 1$, and enhanced power on Lyman-$\alpha$ scales.

\item As a consequence of this mode, our modified perturbation equations exhibit an effective anisotropic stress component. With certain assumptions
about the form of this component on super-Hubble scales, we find that the late-time ISW effect can destructively interfere with the primordial
Sachs-Wolfe amplitude, resulting in a power deficit in the CMB on large scales.

\end{itemize}

Looking forward, our modified gravity models make several key predictions that will be tested by near-future experiments. For instance,
because photons do not couple directly to the longitudinal mode, the weak lensing potential is considerably less affected than the
Newtonian potential. Thus a distinguishing prediction is a discrepancy in the value of $\sigma_8$ estimated from the matter power
spectrum and that inferred from either CMB or weak lensing observations. For $r_c\sim  300-600$~Mpc, the difference is a 20-30\%
effect, depending on the redshift of the observation.

The paper is organized as follows. In Sec.~\ref{motiv}, we describe the building blocks of higher-dimensional generalizations to DGP, following~\cite{giaalpha,degrav}. The result is a two-paramater family of massive or resonance gravity theories, with $\alpha$ specifying the form factor, and $r_c$ the Compton wavelength for the graviton.
Motivated by the DGP Friedmann equation, we propose in Sec.~\ref{backcosmo} a generalized Friedmann equation for higher-dimensional theories.
We then turn to cosmological perturbations in Sec.~\ref{cosmopert}, and motivate our fiducial input parameters for PPF.
Using these, we derive the resulting CMB power spectrum and angular correlation function in Sec.~\ref{cancel}. In Sec.~\ref{strucform}, we study the predictions
for structure formation, including the weak lensing power spectrum (Sec.~\ref{WL}), the Integrated Sachs-Wolfe cross-correlation (Sec.~\ref{ISW}), large scale bulk flows (Sec.~\ref{BF}), the Lyman-$\alpha$ forest (Sec.~\ref{lalpha}), and secondary CMB anisotropies from thermal Sunayev-Zeldovich (SZ) effect (Sec.~\ref{CBISZ}).
We conclude with a summary of our results in Sec.~\ref{conc}.

\section{Phenomenology of higher-dimensional gravity}
\label{motiv}

We are interested in generalizing the phenomenology of the DGP model~\cite{DGP}, arguably the simplest example of modified gravity in the infrared.
In this scenario, our visible universe is confined to a 3-brane in a 4+1-dimensional bulk. Although the bulk has infinite volume, 4$D$ gravity is recovered at short distances, due to
an induced gravity term on the brane. The gravitational force law therefore scales as the usual $1/r^2$ at short distances, but weakens to $1/r^3$ at large distances, with the cross-over scale $r_c$ set by the bulk and brane Planck masses: $r_c = M_4^2/M_5^3$.

In momentum space, this corresponds to the graviton propagator $1/(k^2+r_c^{-1}k)$. Following~\cite{giaalpha,degrav}, we will consider the following power-law generalizations:
\beq
\frac{1}{k^2 + r_c^{-2(1-\alpha)}k^{2\alpha}}\,,
\label{prop}
\eeq
with the DGP model described by $\alpha=1/2$. More generally, $\alpha$ is a freely-specifiable parameter, modulo two constraints: Firstly, in order for the modification to be relevant in the infrared, we must demand that $\alpha < 1$. Secondly, it is straightforward to show that this propagator has a well-defined spectral representation,
and is therefore ghost-free, only if $\alpha$ is positive definite~\cite{giaalpha}. Thus we have the allowed range:
\beq
0\leq \alpha < 1\,.
\label{alprange}
\eeq

The above parametrization is not only phenomenologically simple, but it also makes contacts with DGP ($\alpha=1/2$), as well as its higher-codimension generalizations~\cite{sergei,gigashif,oriol,us}. Since the bulk has $D\geq 6$ space-time dimensions in the latter class of theories, the force law scales as $1/r^{D-2}$ in the infrared,
and the corresponding propagator tends to a constant ($D>6$) or behaves as $\log k$ ($D=6$) as $k\rightarrow 0$. Thus all higher-dimensional extensions
of DGP correspond to $\alpha\approx 0$ theories~\cite{us}.

\subsection{Strong coupling of the longitudinal mode}
\label{moon}

The propagator~(\ref{prop}) describes a resonance, or a continuum of massive states, which is peaked at the tiny scale $r_c^{-1}$.
An immediate consequence is that the graviton now propagates 5 degrees of freedom: the 2 helicity-2 states of Einstein gravity,
2 helicity-1 states, and 1 helicity-0 or longitudinal mode, usually denoted by $\pi$.

On top of the massive spin-2 representation, higher-codimension extensions of DGP also have $D-5$ scalar fields that couple to $T^{\mu}_{\;\mu}$.
For simplicity, we will assume that these scalars have the same propagator as that of the massive graviton~(\ref{prop}),
although this is generally not the case in consistent realizations~\cite{oriol,nonFP}. In other words, for our purposes we will not distinguish between the helicity-0 mode
and the extra scalars, and will denote them collectively as $\pi$.

These extra scalar degrees of freedom contribute to the trace part of the one-particle exchange amplitude~\cite{us}:
\beq
{\cal A} \sim \frac{1}{k^2+ r_c^{-2(1-\alpha)}k^{2\alpha}}\left(T^{\mu\nu}T'_{\mu\nu} - \frac{1}{D-2}TT'\right)\,.
\label{amp}
\eeq
Because the resulting $1/(D-2)$ coefficient differs from the usual $1/2$ of General Relativity (GR), this linearized amplitude is apparently inconsistent with solar system tests. And, remarkably, this departure from GR persists even in the massless limit, $r_c\rightarrow\infty$, a phenomenon famously known as the vDVZ discontinuity~\cite{vDVZ}.

As conjectured by Vainshtein in the context of massive gravity~\cite{vainshtein}, however, these theories can be phenomenologically viable because the linearized approximation for $\pi$ breaks down in the vicinity of astrophysical sources~\cite{ddgv,strong,luty,nicolis}. Below a macroscopic scale $r_\star$, non-linear interactions in $\pi$ become important and result in its decoupling. Power counting arguments~\cite{giaalpha} give
\beq
r_\star = \left(r_c^{4(1-\alpha)}r_{\rm Sch}\right)^{1/(1+4(1-\alpha))}\,,
\eeq
where $r_{\rm Sch}$ is the Schwarzschild radius of the source. And since $r_\star\rightarrow \infty$ as $r_c\rightarrow \infty$, GR is recovered everywhere in this limit, as it should.
Approximate explicit solutions~\cite{ddgv,gruz,por} in DGP have confirmed the Vainshtein effect and the recovery of GR for $r\ll r_\star$.

The leading correction to GR near a source can be inferred as follows. First of all, the solution for $\pi(r)$ must be of order $r_{\rm Sch}/r_\star$ at $r=r_\star$, in order to match the linear solution. Furthermore, within $r_\star$ we expect the solution to be analytic in $r_c^{-2(1-\alpha)}$. These requirements together fix the parametric dependence of $\pi$ for $r\ll r_\star$~\cite{giaalpha,degrav}:
\beq
\pi \sim \frac{r_{\rm Sch}}{r_\star}\left(\frac{r}{r_\star}\right)^{-\frac{1}{2} +2(1-\alpha)}\,.
\eeq
In other words, the correction to the Newtonian potential is of order
\beq
\frac{\delta\Psi}{\Psi} \sim \sqrt{\frac{r}{r_{\rm Sch}}}\left(\frac{r}{r_c}\right)^{2(1-\alpha)}\,.
\label{modpot}
\eeq

An important constraint on such modification comes from Lunar Laser Ranging experiments~\cite{dgz,degrav}:
$\delta\Psi/\Psi < 2.4\times 10^{-11}$~\cite{llr}. Substituting  $r_{\rm Sch} = 0.886\;{\rm cm}$ for the Earth and $r=3.84\times10^{10}\;{\rm cm}$
for the Earth-Moon distance gives
\beq
r_c \gsim 4\cdot 10^{{2(9\alpha-5)\over 1-\alpha}} H_0^{-1}\,.
\label{rclunar}
\eeq
This constraint on $(\alpha,r_c)$ is shown in Fig.~\ref{rc_alpha}. Explicitly, for standard DGP ($\alpha=1/2$) this gives $r_c \gsim 120$~Mpc. For $\alpha=0$, on the other hand,
the bound is far weaker: $r_c\gsim 1$~pc. Future lunar precession experiments will improve these bound by an order of magnitude~\cite{llr}. Somewhat weaker constraints on DGP have also been obtained by studying the effect on planetary orbits~\cite{battat}, and it will be very interesting to extend these studies to general $\alpha$ theories.

\section{Background Cosmology}
\label{backcosmo}

The standard DGP model has a modified Friedmann equation of the form~\cite{cedric}
\beq
H^2 = \frac{8\pi G}{3}\rho \mp \frac{H}{r_c}\,,
\label{friedDGP}
\eeq
where we have assumed spatial flatness. Much attention has been paid to the ``plus" (or self-accelerating) branch of this equation, because of its asymptotically de Sitter solution even in the absence of vacuum energy. However, various arguments have by now established that this branch suffers from ghost-like instabilities~\cite{luty,nicolis,koyamaghost,ruth,dw,rob}. We shall henceforth focus on the ``minus" or healthy branch. Since the $1/Hr_c$ corrections slows down the expansion rate in this case, we must include a cosmological constant to account for cosmic acceleration.

A simple phenomenological extension of this equation suggests itself: $H^2 = 8\pi G\rho/3 - H^{2-\gamma}/r_c^\gamma$. (The ``plus" branch version of this equation was introduced in~\cite{turner} to study generalized self-accelerated solutions.) We can fix $\gamma$ in terms of $\alpha$ as follows. First, from~(\ref{friedDGP}) we see that $\gamma =1$ for standard DGP, corresponding to $\alpha = 1/2$. Furthermore, note from~(\ref{prop}) that $\alpha=1$ can be absorbed into a redefinition of $G$; if this is also the case for the Friedmann equation, then $\gamma = 0$ for $\alpha = 1$. Assuming for simplicity a linear dependence of $\gamma$ on $\alpha$, we conclude that $\gamma = 2(1-\alpha)$, or
\beq
H^2 = \frac{8\pi G}{3}\rho + \Lambda - \frac{H^{2\alpha}}{r_c^{2(1-\alpha)}}\,,
\label{modfried}
\eeq
where $\Lambda$ is a cosmological term. In the higher-dimensional framework, $\Lambda$ should find a natural origin, for instance as a
remnant vacuum energy component from the degravitation mechanism~\cite{degrav}.

Remarkably, this parametrization implies that $\alpha=0$ (corresponding to 2 or more extra dimensions) {\it leads to identical expansion history as $\Lambda$CDM cosmology}.
Of course, higher-codimension DGP models will lead to small deviations from $\Lambda$CDM expansion history, as they do not exactly correspond to $\alpha = 0$. But $\alpha = 0$ should offer a realistic approximation to their modified Friedmann equation since we expect the correction to be a slowly-varying function of $H/r_c$~\cite{claudiaandrew}.

For small values of $\alpha$, we can quantify the expansion history in terms of the usual dark energy parameters $w,w_a,...$:
\bea
\nonumber
\rho_{\rm DE,eff}  &\equiv & \frac{3}{8\pi G}\left(\Lambda - \frac{H^{2\alpha}}{r_c^{2(1-\alpha)}}\right)\,, \\
\nonumber
w & \equiv & -1 - \frac{1}{3}\frac{{\rm d}\ln\rho_{\rm DE,eff}}{{\rm d}\ln a}  \\
\nonumber
&=& -1 -\frac{2\alpha\Omega_{\rm m}+ {\cal O}(\alpha^2)}{(H r_c)^2(1-\Omega_{\rm m})}\,, \\
w_a &\equiv & \frac{{\rm d}w}{{\rm d}\ln a} = \frac{6\alpha\Omega^2_{\rm m} + {\cal O}(\alpha^2)}{(H r_c)^2(1-\Omega_{\rm m})}\,.
\eea
Note that since $\alpha$ is positive, the modified gravity correction leads to an effective dark energy component with $w<-1$. This is indeed the case in the DGP model~\cite{w<-1}. While such phantom behavior may at first seem surprising, it is in fact a natural consequence of the scalar-tensor nature of the effective theory in DGP. Indeed, at the 4$D$ effective level the Friedmann equation is cast in Jordan frame, which can lead to an effective $w<-1$~\cite{DMDE1,DMDE2,staro}.

The current (mainly) geometrical constraint $w = -0.972 \pm 0.060$ (1$\sigma$ error), from a combination of WMAP5, baryonic acoustic oscillations, and supernovae~\cite{Komatsu}, implies $1+w > -0.067$ at 95\% confidence level if we impose the prior that $w\leq -1$, which leads to:
\beq
r_c > 3.2\;\alpha^{1/2}H_0^{-1} \left[1+{\cal O}(\alpha^{2})\right]\,.
\label{rcmax_wmap}
\eeq
In other words, unless $\alpha \lesssim 0.1$, $r_c$ should be larger than today's Hubble radius.

\begin{figure}
\includegraphics[width=\columnwidth]{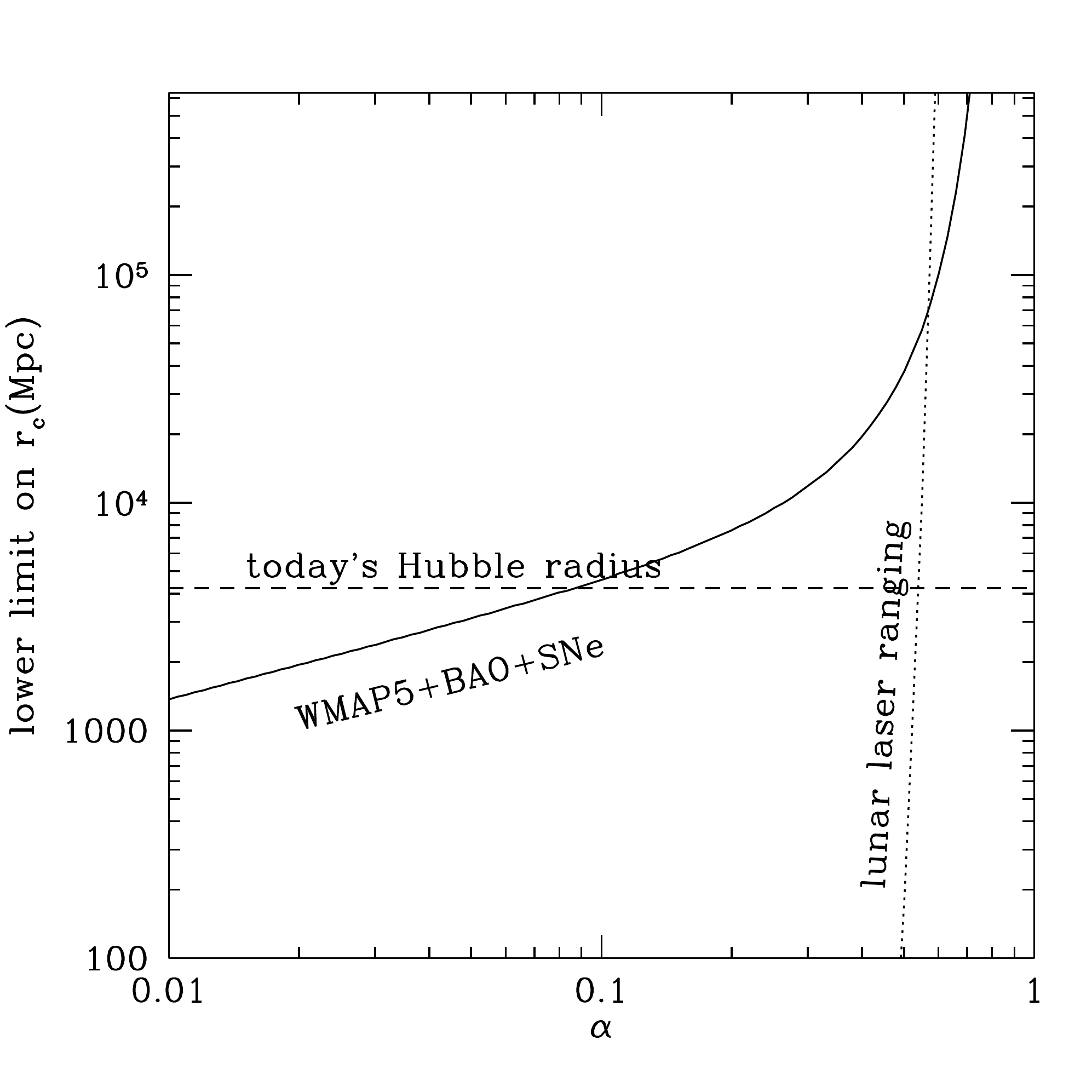}
\caption{95\% lower limits on $r_c$, the scale of transition to higher dimensional gravity, from background cosmology~(see~(\ref{rcmax_wmap})) and lunar laser ranging~(see~(\ref{rclunar})), as a function of $\alpha$. Standard DGP corresponds to $\alpha=1/2$, whereas its higher-dimensional extensions all have $\alpha\approx 0$.}
\label{rc_alpha}
\end{figure}

Figure~\ref{rc_alpha} compares the solar system constraint on $r_c$ --- see~(\ref{rclunar}) --- with the cosmological constraint, if we interpret~(\ref{rcmax_wmap}) as a constraint on the distance to the last scattering surface (assuming a spatially flat universe). We see that for $\alpha \lesssim 0.5$, the main constraint on the value of $r_c$ comes from the background cosmology. However, as we noted above,  in the $\alpha \rightarrow 0$ limit (corresponding to 2 or more extra dimensions), the background cosmology is indistinguishable from $\Lambda$CDM, and thus $r_c$ remains unconstrained. We will see in the next section that cosmological perturbations will be sensitive to $r_c$ even in this limit.

\section{Cosmological perturbations}
\label{cosmopert}

Since the study of cosmological perturbations in explicit higher-codimension extensions of DGP has yet to be worked out, for our analysis we must resort
to a phenomenological parametrization of these modifications. In this paper we take  the parametrized post-Friedmann (PPF) approach introduced recently~\cite{ppf}. This formalism is reviewed in Appendix \ref{ppf}.

In Newtonian gauge, scalar metric perturbations are specified by the gauge-invariant potentials $\Psi$ and $\Phi$:
\beq
ds^2 = -(1+2\Psi){\rm d}t^2 + a^2(1+2\Phi){\rm d}\vec{x}^2\,.
\eeq
In the absence of anisotropic stress, the momentum constraint in GR fixes $\Phi = -\Psi$. However, this is no longer true in DGP because of the helicity-0 mode.
At the 4D effective level, this can be understood tracing back to the scalar-tensor nature of the theory, as it is well-known that $\Phi\neq \Psi$ in
Jordan frame. This deviation from GR is parametrized in PPF through
\beq
g(a,k) \equiv \frac{\Phi+\Psi}{\Phi-\Psi}\,.
\label{gmain}
\eeq
This function can be specified independently from the expansion history.

At this point we should mention closely related parametrizations of modified gravity that have appeared in the literature.
Jain and Zhang~\cite{bhuvnesh}  parametrized the deviation from $\Phi = -\Psi$ through $\eta = 1+g/(1-g)$. This convention
was also adopted by Bertschinger and Zukin~\cite{ed2}, except that the parameter was denoted by $\gamma$.
Daniel {\it et al.}~\cite{caldwell}, on the other hand,  introduced $\varpi = -2g/(1+g)$. A key difference is that
these parameters were all assumed to depend on $a$ only, whereas our $g(a,k)$ will also be a function of scale.

For future reference, it is convenient to introduce the lensing potential
\beq
\Phi_- \equiv \frac{\Phi-\Psi}{2}\,.
\label{Phi-}
\eeq
Gravitational lensing observations, as well as the Integrated Sachs-Wolfe (ISW) effect, are both sensitive to $\Phi_-$. Galaxy peculiar velocity
measurements, on the other hand, are determined by the Newtonian potential $\Psi= (g-1)\Phi_-$. See~\cite{bhuvnesh} for a pedagogical description of
how various observables relate to different metric combinations.

In the next few subsections, we will specify $g(a,k)$ for our modified gravity theories, based on known results in standard DGP.
An important lesson from studies of DGP is that $g(a,k)$ has qualitatively different behaviors on sub- or super-horizon scales. Hence
we will explore these two regimes separately.

\subsection{Sub-horizon evolution}

Focusing on sub-Hubble scales and non-relativistic sources in standard DGP, Lue {\it et al.}~\cite{lue} obtained
\beq
\left.g_{\rm DGP}\right\vert_{k\gg aH} = -\frac{1}{3}\cdot\frac{1}{\left[1 + 2 Hr_c\left(1+\frac{H'}{3H}\right)\right]}\,,
\label{gDGP}
\eeq
where $'\equiv d/d\ln a$. Note the plus sign within the brackets, which is appropriate for the normal branch~\cite{song}.
Apart from this sign, this result was also obtained in cosmological perturbation theory using a quasi-static approximation~\cite{koyama}.

The dependence on $Hr_c$ in~(\ref{gDGP}) can be attributed to the dynamics of $\pi$. Indeed, for the universe as a whole, both $r$ and $r_{\rm Sch}$ can be
approximated by $H^{-1}$. Setting $\alpha=1/2$ in this case,~(\ref{modpot}) gives the leading correction to standard cosmology,
$\left.\delta\Psi/\Psi\right\vert_{\rm DGP} \sim 1/Hr_c$, in agreement with~(\ref{gDGP}) in the strong coupling regime $Hr_c\gg 1$.
Repeating the argument for general $\alpha$,~(\ref{modpot}) gives
\beq
\frac{\delta\Psi}{\Psi} \sim \frac{1}{(Hr_c)^{2(1-\alpha)}}\,.
\eeq
Meanwhile, in the weak coupling regime, from the exchange amplitude~(\ref{amp})
we can read off $\delta\Psi/\Psi = -(D-4)/(D-2)$, consistent with the numerical coefficient of~(\ref{gDGP}) for $D=5$.
Taking into account this numerical coefficient, this suggests generalizing~(\ref{gDGP}) to
\beq
\left.g\right\vert_{k\gg aH} = -\left(\frac{D-4}{D-2}\right)\frac{1}{1 + 2 \left(Hr_c\right)^{2(1-\alpha)}\left(1+\frac{H'}{3H}\right)}\,.
\label{g>}
\eeq
%
A similar parametrization of $g$ for general $\alpha$ theories was also studied in~\cite{koyamaparam}, albeit for self-accelerated cosmologies.
Our parametrization is also related to the slip parameter $\varpi$ introduced in~\cite{caldwell}. The precise translation, as mentioned earlier, is
$\varpi = -2g/(1+g)$. In the matter-dominated era, these authors assumed $\varpi \sim 1/H^2$, which coincides with the scaling of $g$
in the case $\alpha = 0$.

\subsection{Super-horizon evolution and decay of Newtonian potential}

In standard DGP, the evolution of super-horizon fluctuations was studied in~\cite{hu} using a scaling ansatz for bulk perturbations.
On the self-accelerating branch, the solution is well-fitted by~\cite{ppf}
\beq
\left.g_{\rm DGP}\right\vert_{k\ll aH} = \frac{9}{8Hr_c-1}\left(1+\frac{0.51}{Hr_c-1.08}\right)\,.
\label{gDGP2}
\eeq
The above parametric dependence and the form of our sub-Hubble parametrization~(\ref{g>}) suggest the following ansatz for general $\alpha$,
\beq
\left.g \right\vert_{k\ll aH} = \frac{A}{1 + 2 \left(Hr_c\right)^{2(1-\alpha)}\left(1+\frac{H'}{3H}\right)}\,.
\eeq
The normalization constant $A$ will be fixed shortly.

Using a simple interpolation between the self-accelerated analogue of the short-wavelength $g_{\rm DGP}$ given in~(\ref{gDGP}) and its long-wavelength counterpart~(\ref{gDGP2}),
Wang {\it et al.}~\cite{wiley} recently showed that the self-accelerated DGP model is disfavored at the $\sim 5\sigma$ level compared to $\Lambda$CDM. Their analysis includes
CMB temperature anisotropy, supernovae data, and constraints on the Hubble constant. Roughly 70\% of the tension comes from the modification in the Friedmann equation, while
the remaining 30\% is due to the difference in growth history, resulting in too large an ISW contribution to the low-$\ell$ multipoles of the CMB.

Similarly, we find that an unacceptably large ISW component is a generic outcome for general $\alpha$ theories, unless $r_c$ is much larger than the present Hubble radius.
But there is one important and very interesting exception: by choosing $A\approx 1$ and making a few minor assumptions on the remaining PPF parameters to be discussed shortly, the
Integrated Sachs-Wolfe effect resulting from the modified growth history can cancel the primordial Sachs-Wolfe contribution, resulting in a power deficit on large angular scales. This novel mechanism
{\it can therefore naturally account for the observed low quadrupole as well as the lack of angular correlations above} $\sim 60\,^{\circ}$~\cite{cobe,wmap1,huterer,others,huterer2}. See Fig.~\ref{cmb}.

The actual value of $A$ in explicit higher-codimension modifications of gravity remains at present unknown and, in any case, is likely to be model-dependent. Thus we view
the agreement with large-angular CMB anomalies for $A=1$ as a strong observational guide towards building successful models of infrared modified gravity.
The onus is then on model-builders to find explicit brane-world constructions where $A=1$ is realized. For present purposes, we shall assume this can be achieved
and explore the consequences for other observables. Let us therefore take
\beq
\left.g \right\vert_{k\ll aH} = \frac{1}{1 + 2 \left(Hr_c\right)^{2(1-\alpha)}\left(1+\frac{H'}{3H}\right)}\,.
\label{g<}
\eeq
Note that the transition $g\rightarrow 1$ has a natural physical interpretation: from the definition of $g$ in~(\ref{gmain}), this corresponds
to {\it a decay of the Newtonian potential $\Psi$ relative to the lensing potential $\Phi_-$}.

At this point the astute reader might worry about the sign of $g\vert_{k\ll aH}$. On sub-Hubble scales, a positive $g$ would be interpreted as $\pi$ mediating a repulsive force
and therefore representing a ghost. Fortunately, such reasoning does not apply on super-horizon scales, and it is straightforward to devise toy scalar-tensor theories in $3+1$
dimensions that lead to a positive $g$ on super-Hubble scales. For instance, as shown in Appendix \ref{Brans-Dicke}, all Brans-Dicke theories satisfy this property during the
matter-dominated era.

\subsection{Summary of fiducial model}
\label{fiducial}

The full $g(a,k)$ must interpolate between the sub- and super-Hubble behavior given in~(\ref{g>}) and~(\ref{g<}), respectively. For this purpose, we choose
the interpolation form
\bwt
\beq
g(a,k) = \frac{1}{1 + 2 \left(Hr_c\right)^{2(1-\alpha)}\left(1+\frac{H'}{3H}\right)}\left[\frac{1}{1+\left(c_g\frac{k}{aH}\right)^a}
- \left(\frac{D-4}{D-2}\right)\frac{\left(c_g\frac{k}{aH}\right)^b}{1+\left(c_g\frac{k}{aH}\right)^b}\right]\,,
\label{ginterpol}
\eeq
\ewt
where $a,b$ and $c_g$ are constants. While our results are not overly sensitive to the precise values for these constant parameters, in practice
we have found that smoother interpolations offer a better fit to the low multipoles of the CMB power spectrum. Hence, as fiducial values we take
\beq
a = b = 0.5\; ; \qquad c_g = 0.5\,.
\eeq
Although the interpolation is smooth, one can think of $c_g$ as setting the scale below which $g$ becomes negative. Note that for the above choice
of $c_g$, this scale is comparable to the horizon. Larger values of $a,b$ and $c_g$ diminish the effectiveness of the cancelation mechanism.

Another key input parameter in PPF is $c_\Gamma$, which pertains to the evolution of the curvature perturbation $\zeta$ (see~(\ref{zetadef}) for its definition).
Independent of the theory of gravity under consideration, energy-momentum conservation alone imposes that $\zeta$ is conserved in the infinite-wavelength
limit~\cite{separate}. Using this fact, Bertschinger~\cite{ed} derived the following consistency relation
\bea
\nonumber
& & \zeta = {\rm const.} = \frac{H}{H'}\left[(g-1)\Phi_- -g'\Phi_- - (g+1)\Phi_-'\right] \\
& & \;\;\;\;\;\;\;\;\;\;\;\;\;\;\;\;\;\;\;\;\;\;\;\;\;\;\;+ (g+1)\Phi_-\,.
\label{zetacons}
\eea
In general, this condition must hold well outside the horizon. As we will see shortly, however, here we impose that $\zeta$ is conserved on all scales in our fiducial
modified gravity model.

While $\zeta$ is actually conserved on all scales in $\Lambda$CDM cosmology, this is not necessarily the case for modified gravity theories.
In the PPF approach, one specifies a parameter $c_\Gamma$ which, as described in Appendix \ref{ppf}, determines an effective horizon
$c_\Gamma/aH$ above which $\zeta$ is conserved. Our mechanism for canceling the large-scale CMB power requires that conservation of $\zeta$
persist on a sufficiently wide range of sub-Hubble modes, corresponding to small values of $c_\Gamma$. In practice, we have found that
$c_\Gamma\;\lsim\; 10^{-2}$ is desirable. For simplicity, however, as fiducial value we shall set
\beq
c_\Gamma = 0\,,
\eeq
thereby enforcing $\zeta$ conservation {\it on all scales}. The evolution of super-horizon modes is therefore uniquely determined through~(\ref{zetacons})
by specifying $g$ and the expansion history. Note that the assumption of constant $\zeta$ was also made by Bertschinger and Zukin~\cite{ed2} in what
they dubbed scale-independent modified gravity models. (A key difference, however, is that their $g$ was also independent of scale.)
In Appendix~\ref{ppf}, we shall describe the PPF formalism in more generality, allowing for non-zero $c_\Gamma$.

\begin{figure}
\includegraphics[width=\columnwidth]{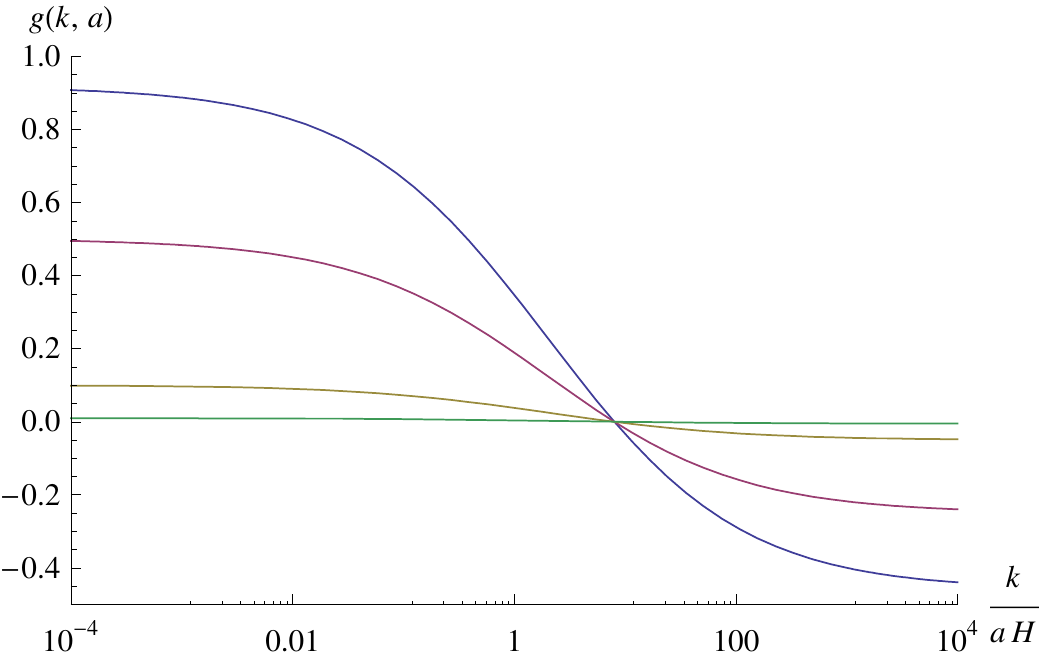}
\caption{Dependence of $g(k,a)$ (see~(\ref{ginterpol})) in our model on $k/aH$ in the matter-dominated era for $r_c = 0.3,1,3$, and $10$ Hubble radius. The number of space-time dimensions, $D$, is assumed to be $6$ (i.e. $2$ extra dimensions). General Relativity is recovered when $r_c \gg H^{-1}$ as, $g\rightarrow 0$.}
\label{g_fig}
\end{figure}
In light of the relatively large number of input functions and parameters inherent to the PPF framework, let us synthesize the
essential qualitative and quantitative features of our fiducial parametrization:

\begin{itemize}

\item The difference in metric potentials $g(a,k)$ is chosen to interpolate between the sub-horizon behavior, dictated by well-understood features
of massive/resonance gravity outlined in Sec.~\ref{moon}, and the more model-dependent super-horizon behavior.
For the latter, the large-scale CMB anisotropy constrains us to choose $A=1$, so that $g|_{k\ll aH}$ makes a transition from
0 at early times to 1 at late times --- see Fig. \ref{g_fig}. Physically, this corresponds to a decay of the Newtonian potential $\Psi$ relative to $\Phi$.
Smoother interpolations are preferable.

\item The curvature perturbation $\zeta$ must be nearly conserved, which requires small values for $c_\Gamma$. For simplicity, we set $c_\Gamma = 0$ in our fiducial model.

\end{itemize}

The remaining parameters to vary are of course $\alpha$, $D$ and $r_c$. Since we are interested in higher dimensional generalizations of DGP, we shall fix
\beq
\alpha = 0\,,
\eeq
as discussed below~(\ref{alprange}).
Since the expansion history reduces to $\Lambda$CDM in this case, we are free to consider values of $r_c$ that are relatively small compared to the Hubble radius.
We will find that our modified gravity theories can explain the anomalies listed in the Introduction and match other observations with $r_c \sim 300-600$~Mpc. Finally, the total number of
space-time dimensions $D$ sets the strength of the extra scalar force at short distances, as seen in~(\ref{g>}), with larger $D$ corresponding to stronger gravity.
To explain the CBI excess and the anomalous Lyman-$\alpha$ power, it will suffice to consider two extra dimensions, or
\beq
D = 6\,.
\eeq
This completes the description of our fiducial model.

\section{CMB Temperature Anisotropy and Large-Scale Anomalies}
\label{cancel}

Let us step back and explain why the choice for $g \vert_{k\ll aH}$ in~(\ref{g<}) can explain the low quadrupole and the lack of correlation on $\gsim 60\,^{\circ}$~scales. On large angles, the CMB anisotropy is well approximated by the sum of the Sachs-Wolfe and the Integrated Sachs-Wolfe (ISW) effects \cite{Sachs:1967er}:
\beq
\frac{\delta T_{\rm CMB}}{T_{\rm CMB}} = \frac{1}{3}\Phi_- +2\int {\rm d}t \frac{\partial \Phi_-}{\partial t} \simeq  \frac{1}{3}\Phi_- + 2\Delta \Phi_-\,.
\label{isw_sw}
\eeq
The integral in the ISW effect is taken along the light cone, while the partial derivative is at fixed comoving spatial position. Therefore, the last approximation in~(\ref{isw_sw}) is only valid when the spatial variations of $\Phi_-$ are {\it much smaller} than its time variation: $|\vec{\nabla}\Phi_-| \ll |\dot{\Phi}_-|$.
It is then interesting to note that on these scales, $\Delta \Phi_-$ can be obtained analytically as a function of $g(a,k)$, from the fact that the curvature perturbation $\zeta$ remains
constant on super-horizon scales --- see~(\ref{zetacons}). In particular, if $g$ goes from 0 to a finite (constant) value $g_{\rm f}$ during the matter era, from~(\ref{zetacons}) we see that:
\beq
\Phi_- \rightarrow \left(\frac{5}{5+g_{\rm f}}\right)\Phi_- \Rightarrow \Delta\Phi_- = -\left(\frac{g_{\rm f}}{5+g_{\rm f}}\right)\Phi_-\,\label{Phi_-_ISW}\,.
\eeq
For example, for the transition from 0 to $g_{\rm f}=1$ we have:
\beq
\frac{\delta T_{\rm CMB}}{T_{\rm CMB}} \simeq \frac{1}{3}\Phi_- + 2\Delta \Phi_- = \left[5(1-g_{\rm f})\over 3(5+g_{\rm f})\right] \Phi_- = 0\,!
\eeq
In other words, if $g\rightarrow 1$ at late times in the modified gravity theory, the ISW and Sachs-Wolfe effects exactly cancel on large angles --- corresponding to super-horizon scales at the time of the transition ---, which could potentially explain the vanishing of CMB power on scales $\gtrsim 60^{\circ}$. Note that $g\rightarrow 1$ is equivalent to a vanishing Newtonian potential relative to the lensing potential,
$\Psi \ll \Phi_-$, at late times. A partial cancelation between the primordial Sachs-Wolfe and a late-time ISW contribution arising from $g$ was also noticed in~\cite{ed2,caldwell}.

\begin{figure}
\includegraphics[width=0.99\columnwidth]{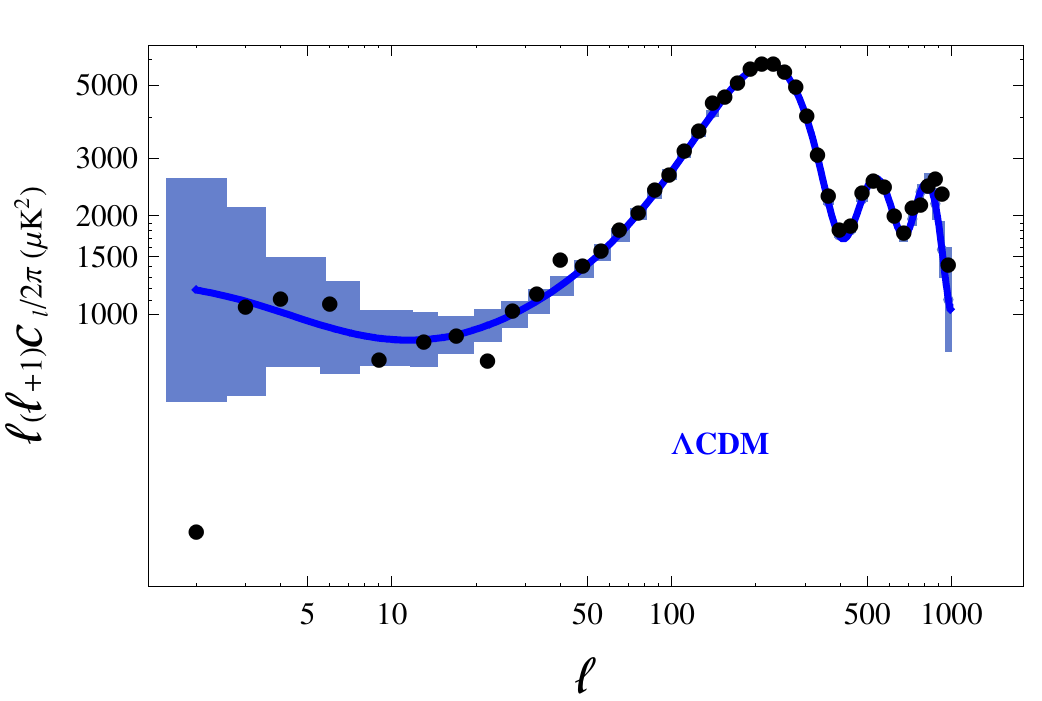}
\includegraphics[width=0.99\columnwidth]{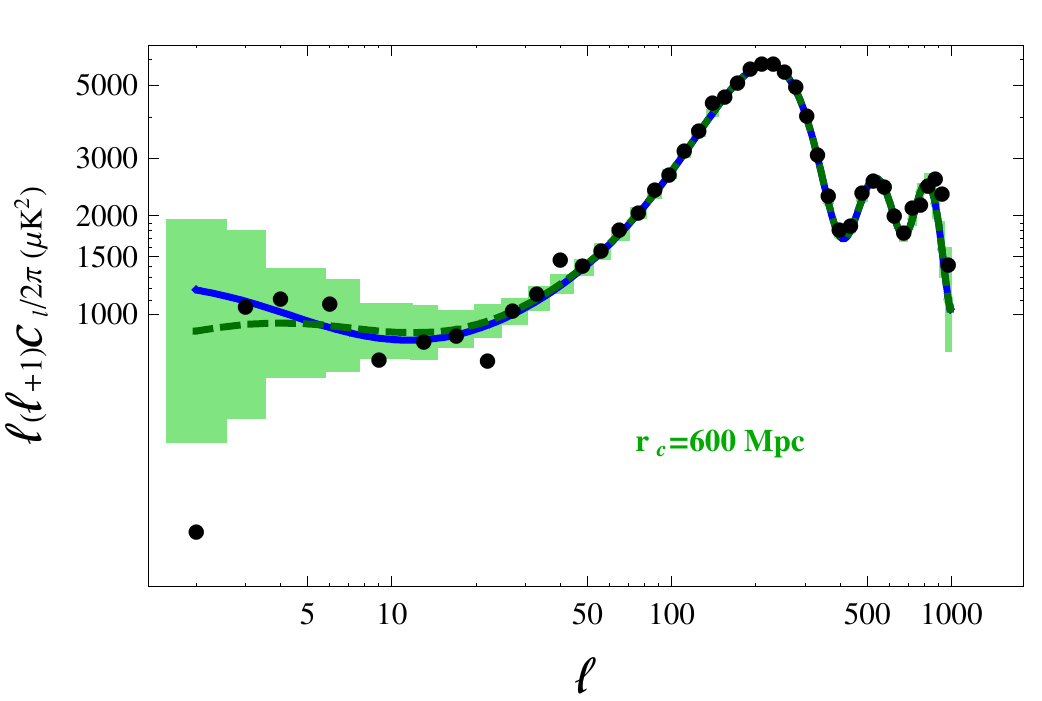}
\includegraphics[width=0.99\columnwidth]{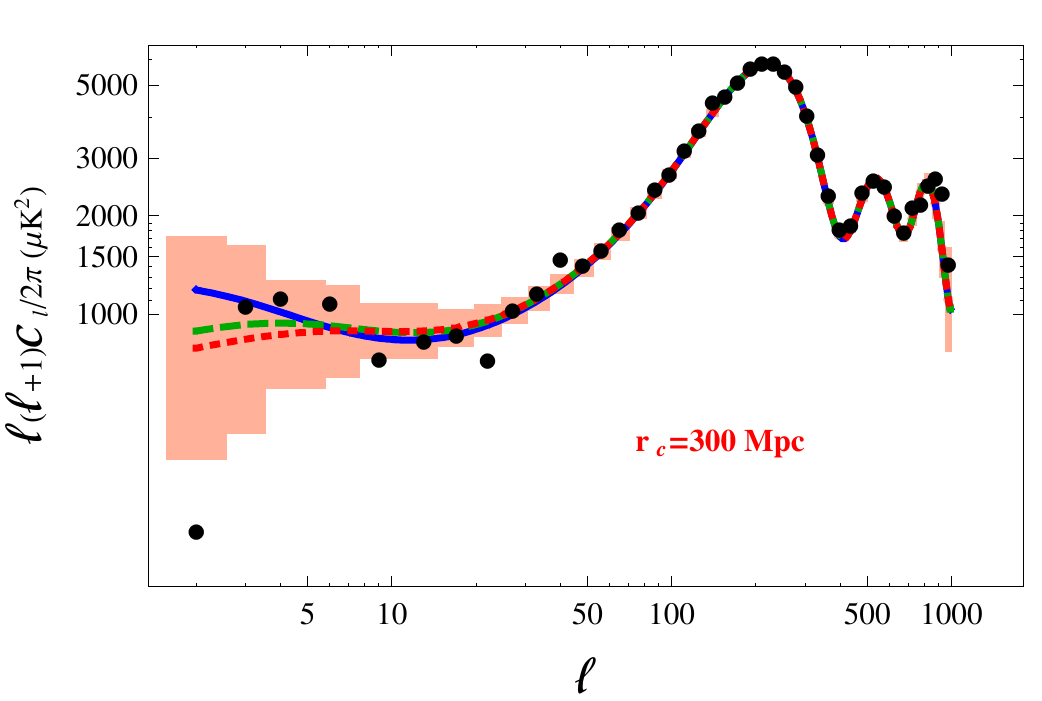}
\caption{CMB angular power spectra for best-fit $\Lambda$CDM (blue, solid curve), $r_c = 600$~Mpc (green, dashed curve) and $r_c = 300$~Mpc (red, short-dashed curve).
The vertical bars show the cosmic variance spread centered on the respective theoretical predictions. In the middle and bottom panels, the previous curves are shown for comparison.
The data points are from the WMAP-5 data release~\cite{Komatsu}.}
\label{cmb}
\end{figure}
\begin{figure}
\includegraphics[width=\columnwidth]{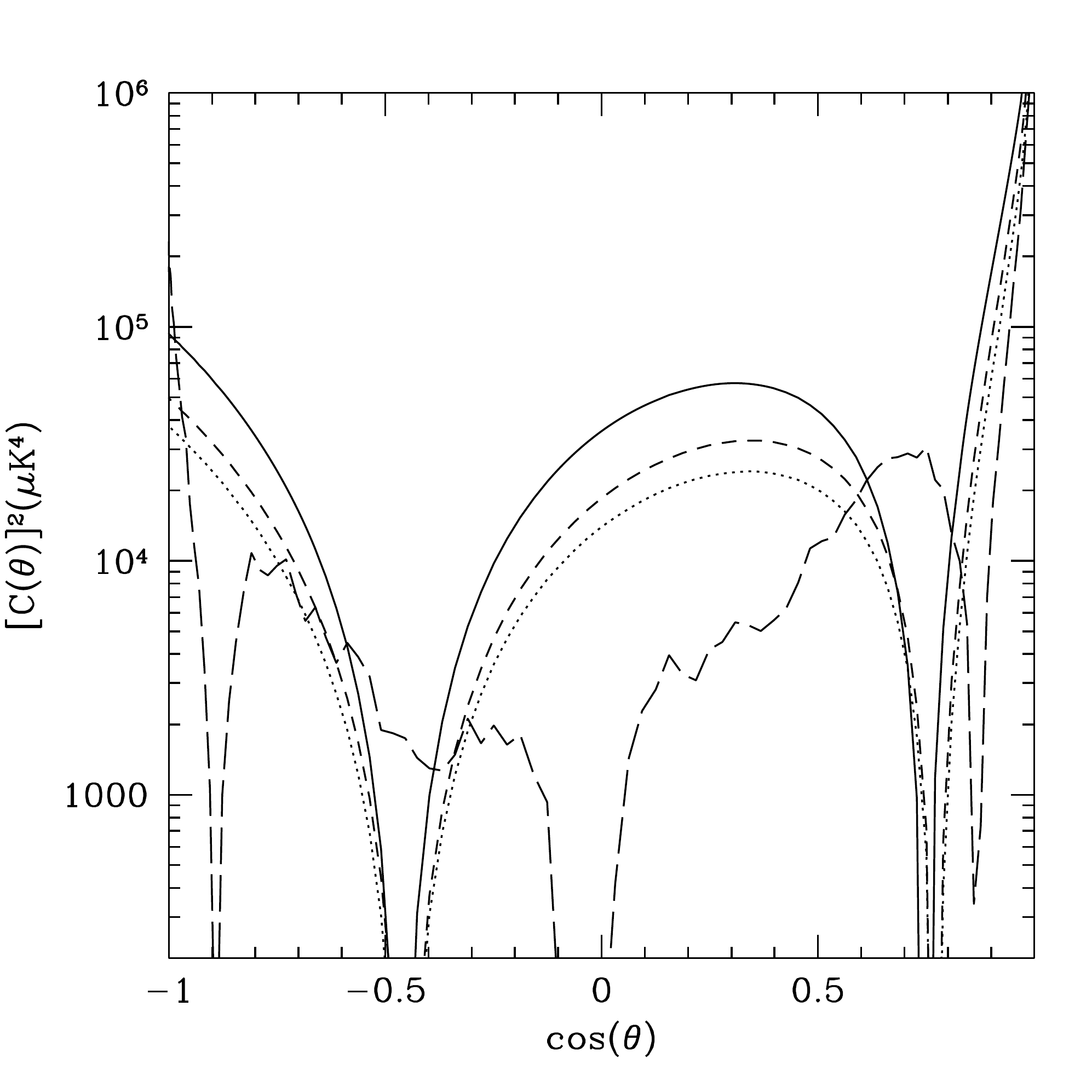}
\caption{Square of the correlation function of CMB temperature anisotropy, as a function of the cosine of the separation angle in the sky. The curves show $\Lambda$CDM (solid), $r_c = 300$ Mpc (dotted) and $r_c = 600$ Mpc (short-dashed). The long-dashed curve is the Legendre transform of the WMAP5 maximum likelihood power spectrum~\cite{Nolta:2008ih}. The observed correlation is systematically below the $\Lambda$CDM prediction for $\theta \gtrsim 60^{\circ}$, or $\cos\theta < 0.5$.}
\label{cmb_corr}
\end{figure}

This phenomenological observation, in addition to the fact that Brans-Dicke theories have $g>0$ on super-horizon scales --- see Appendix~\ref{Brans-Dicke} ---, justifies the infrared limit of our ansatz for $g(a,k)$ given in~(\ref{ginterpol}).

Figure~\ref{cmb} demonstrates how this cancelation can work at low $\ell$'s for different values of $r_c$. The CMB power spectra were generated
using a version of CAMB~\cite{camb} modified by W.~Fang to include PPF parameters~\cite{cambppf}.

\begin{figure}
\includegraphics[width=\columnwidth]{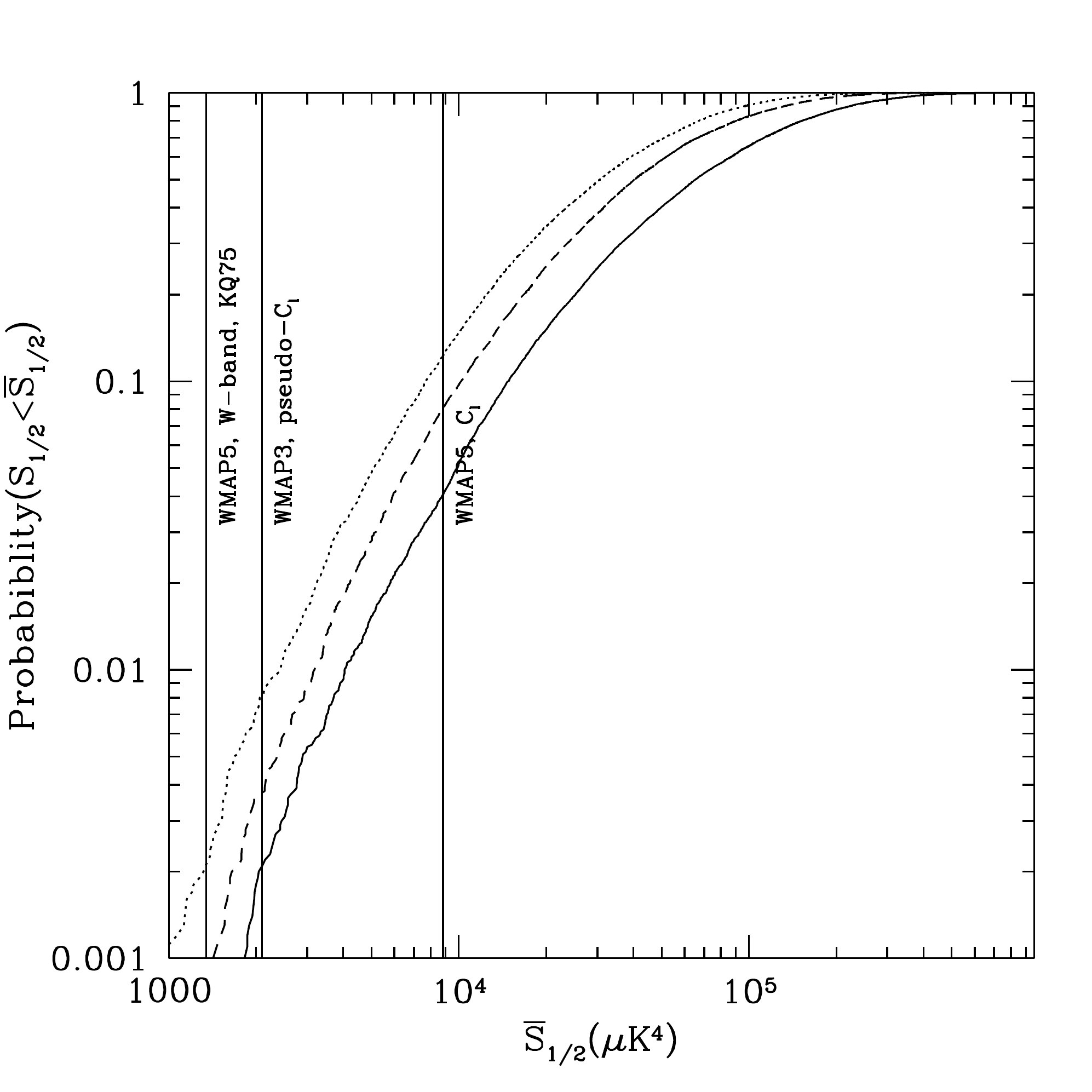}
\caption{The cumulative probability for $S_{1/2}$,  defined in~(\ref{eq:s12}), measuring the CMB correlation on scales $> 60^{\circ}$. The curves are for $\Lambda$CDM (solid), $r_c = 300$ Mpc (dotted) and $r_c = 600$ Mpc (dashed). The vertical lines show the observed values of $S_{1/2}$ from different estimators (see~\cite{huterer2} for details).}
\label{s12_dgp}
\end{figure}

The apparent lack of power on large angles was first quantified by Spergel et al. 2003 in the first data release of WMAP~\cite{Spergel:2003cb}, through the quantity:
\beq
S_{1/2} \equiv \int_{-1}^{1/2} [{\cal C}(\theta)]^2 {\rm d}\cos\theta\,,\label{eq:s12}
\eeq
where ${\cal C}(\theta)$ is the angular correlation function of the CMB temperature anisotropies shown in Fig.~\ref{cmb_corr}. Most recently, Copi {\it et al.}~\cite{huterer, huterer2} argued that the probability of $S_{1/2}$ to be lower than it is observed outside the Galactic plane is only $0.025\%$ in the concordance $\Lambda$CDM model. However, different estimators for the correlation function can lead to larger values of $S_{1/2}$. Figure~\ref{s12_dgp} compares this probability for our modified gravity models with $\Lambda$CDM, and shows that it is increased by about a factor of 2-3 in the relevant range. While this is not enough to explain the low values of $S_{1/2}$ in direct measurements of correlation function [{\it e.g.}, the two smaller values in Fig.~\ref{s12_dgp}], it can bring the prediction within 90\% probability range for maximum likelihood estimates of the WMAP5 power spectrum [the larger value in Fig.~\ref{s12_dgp}].

\begin{figure}
\includegraphics[width=\columnwidth]{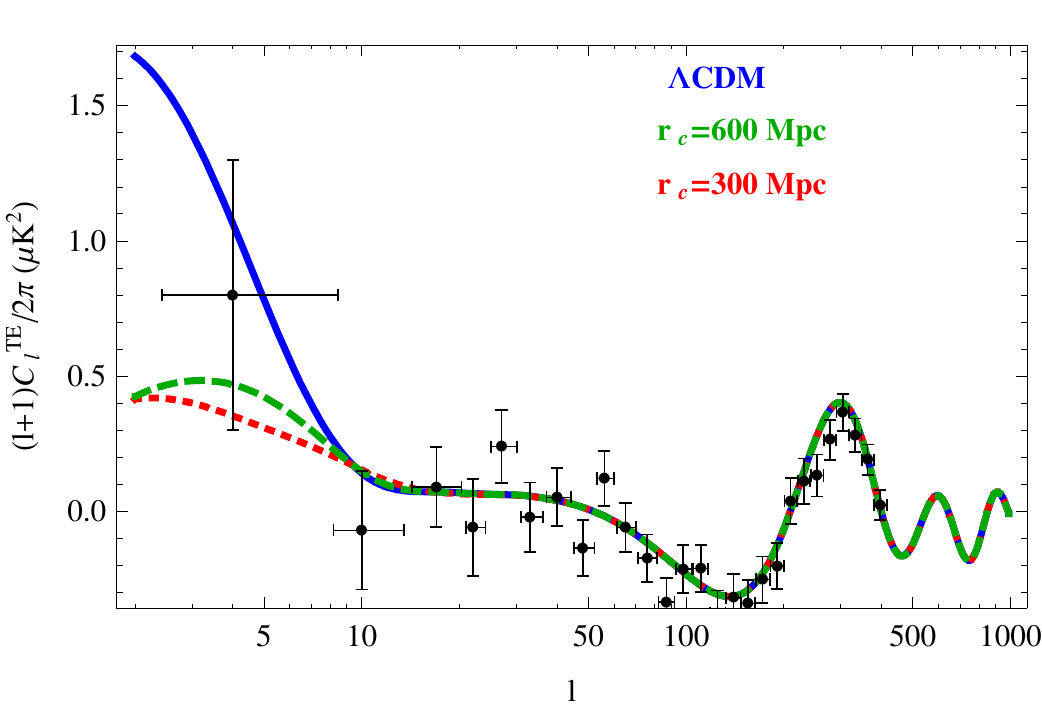}
\includegraphics[width=\columnwidth]{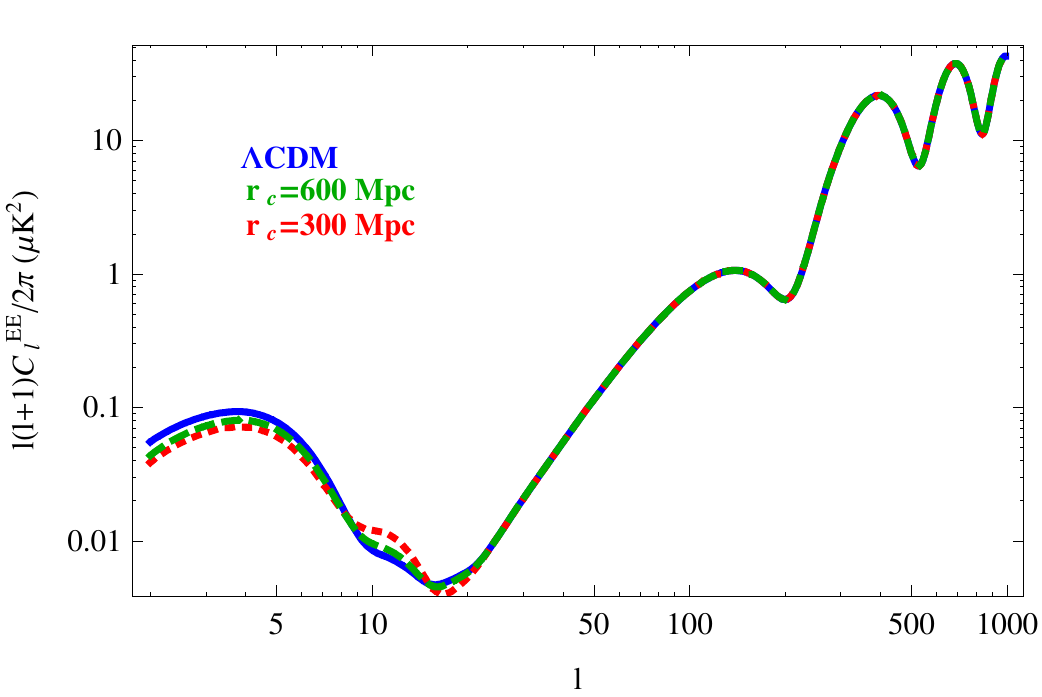}
\caption{Temperature-Polarization (TE) and polarization (EE) power spectra for $\Lambda$CDM (blue, solid curve), $r_c = 300$ Mpc (red, short-dashed curve) and $r_c = 600$ Mpc (green, dashed curve). The points are from the WMAP5 data release~\cite{Nolta:2008ih}.}
\label{polar}
\end{figure}

Despite the improvement of the consistency between models and data at large angles for temperature anisotropy, the $\chi^2$ marginally increases by 1.21 (0.32) for $r_c=$ 300 (600) Mpc, with other cosmological parameters fixed. The origin of this increase can be seen in Fig.~\ref{polar}, where the suppression of power at small $\ell$'s decreases the TE temperature-polarization cross-power spectrum below the observed value. However, given that the lowest $\ell$-bin in the TE spectrum decreased by a factor of $3$ from the 1st year to the 5-year WMAP data release due to an improved foreground cleaning, it is still conceivable that the systematics in this bin are underestimated. Therefore, we predict a {\it significantly lower TE cross-power spectrum} at $\ell <10$, which should be clearly distinguished from $\Lambda$CDM by the {\it Planck} satellite, due to its better polarization sensitivity and foreground cleaning capabilities~\cite{Planck}.

Meanwhile, as shown in the bottom panel of Fig.~\ref{polar}, the EE spectrum is not appreciably affected by our modification of gravity.

We should also point out that it is possible to improve the full $\chi^2$ (relative to $\Lambda$CDM) without extra marginalization, simply by decreasing $b$ in~(\ref{ginterpol}). However, this will spoil our prediction of extra growth for small-scale structures, which we shall discuss next.

\section{Structure Formation with massive gravity}
\label{strucform}

\begin{figure}
\includegraphics[width=\columnwidth]{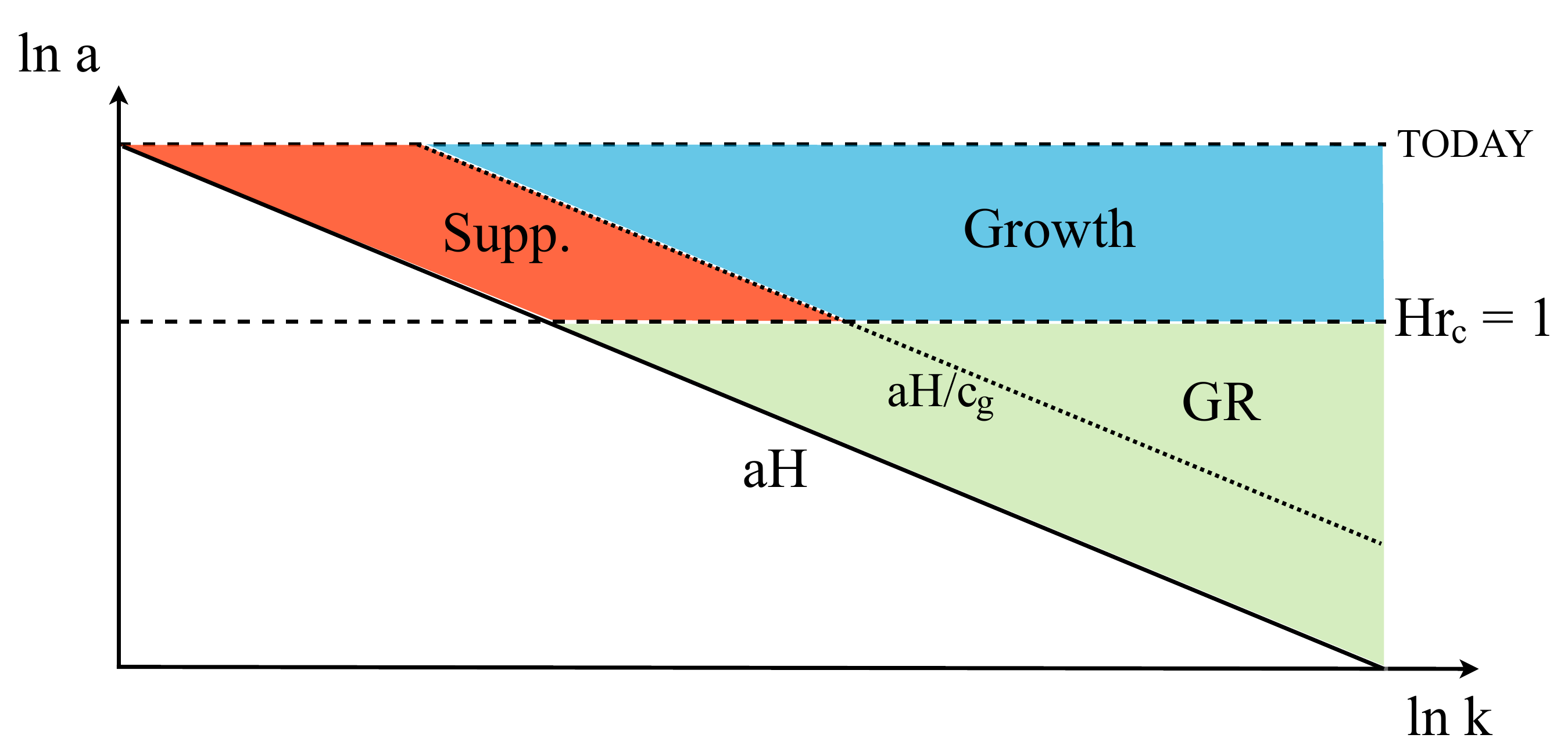}
 \caption{Different growth regimes as function of scale factor and comoving wavenumber $k$. Colored regions correspond
 to modes inside the horizon today. The solid line labeled by $aH$ denotes the Hubble horizon; the dotted line traces the
 characteristic scale set by $c_g$. At early times, $H > r_c^{-1}$, growth proceeds as in GR (Green region); for $H < r_c^{-1}$ and on scales much smaller than
  the $c_g$-scale, growth is enhanced thanks to the helicity-0 mode (Blue region);
for $H < r_c^{-1}$ and on large scales, the decay of the Newtonian potential $\Psi$ suppresses growth (Red region).}
 \label{sketch}
 \end{figure}

The growth of perturbations in our model can be understood analytically by making a few simplifying approximations.
The essential physics is illustrated in Fig.~\ref{sketch}. The solid line labeled by $aH$ denotes
the Hubble horizon. Meanwhile, the dotted line denotes the scale set by $c_g$ where $g$ makes a transition from
positive values on larger scales to negative values on smaller scales.

At early times, $H\gg r_c^{-1}$, we have a GR regime on all scales, as $g \approx 0$.
For modes much smaller than the $c_g$-scale, once $H$ drops below $r_c^{-1}$
the scalar force mediated by the longitudinal mode kicks in, enhancing the growth of perturbations.
This is the blue-shaded region in Fig.~(\ref{sketch}). Thus there is excess of power on small scales, compared to $\Lambda$CDM.

For modes with wavelength above the $c_g$-scale, the function $g$ makes a transition from 0 to 1 at $H\sim r_c^{-1}$, corresponding
to a decay of the Newtonian potential $\Psi$. This is the red-shaded region in Fig.~\ref{sketch}. This suppression of growth results in
a power deficit in the spectrum of density perturbations compared to $\Lambda$CDM.

Before numerically solving for the evolution of perturbations, we can gain analytic insights by studying different regimes.
Although our perturbed metric is cast in Newtonian gauge, we find it more convenient to track
{\it comoving gauge} density perturbation, $\Delta_{\rm m}$. Its evolution equation follows as usual
by combining energy and momentum conservation [see~(\ref{Emomap})]
\beq
\Delta_{\rm m}'' + \left(2+\frac{H'}{H}\right)\Delta_{\rm m}' = (1-g)\frac{k^2}{a^2H^2}\Phi_-\,,
\label{delevo}
\eeq
where we have used the fact that the curvature perturbarion is assumed to be conserved on all scales, $\zeta' = 0$.
As it should,~(\ref{delevo}) reduces to the standard evolution equation in the GR limit $r_c\rightarrow\infty$.

At late times ($H \lsim r_c^{-1}$), when the transition is complete, $g(k,a)$ only varies slowly as the mode wavelength relative to the horizon shrinks with time --- see Fig.~\ref{g_fig}. In this regime we can ignore $\Phi'_-$ and $g'$ in the $\zeta$ equation (\ref{zetacons}), which leads to~(\ref{Phi_-_ISW}) in the matter era. With this in mind, we can view~(\ref{delevo}) as a forced damped harmonic oscillator, where the force is enhanced by a factor of $5(1-g)/(5+g)$ compared to $\Lambda$CDM. Therefore, prior to the domination of $\Lambda$ the inhomogeneous solution becomes
\beq
\Phi_- \approx \frac{5}{5+g}\Phi_-^{\rm CDM}\,;\;\;\;  \Delta_{\rm m} \approx \frac{5(1-g)}{5+g}\Delta_{\rm m}^{\rm CDM}\,.
\eeq
On very small scales, where $g \rightarrow -(D-4)/(D-2)$, this gives
\beq
\Phi_- \rightarrow \frac{5(D-2)}{2(2D-3)}\Phi_-^{\rm CDM} \;;\;\;  \Delta_{\rm m} \rightarrow \frac{5(D-3)}{2D-3}\Delta_{\rm m}^{\rm CDM}\,.
\eeq
In particular, for our fiducial case of $D=6$,
\beq
\Phi_- \rightarrow \frac{10}{9}\Phi_-^{\rm CDM} \;;\;\;  \Delta_{\rm m} \rightarrow \frac{5}{3}\Delta_{\rm m}^{\rm CDM}\,.
\label{Phi109}
\eeq
Meanwhile, on large scales we have $g \rightarrow 1$, and thus $\Delta_{\rm m} \rightarrow 0$ while $\Phi_-$ decreases by a factor of $5/6$, as we discussed in the previous section.

\begin{figure*}
\includegraphics[width=\columnwidth]{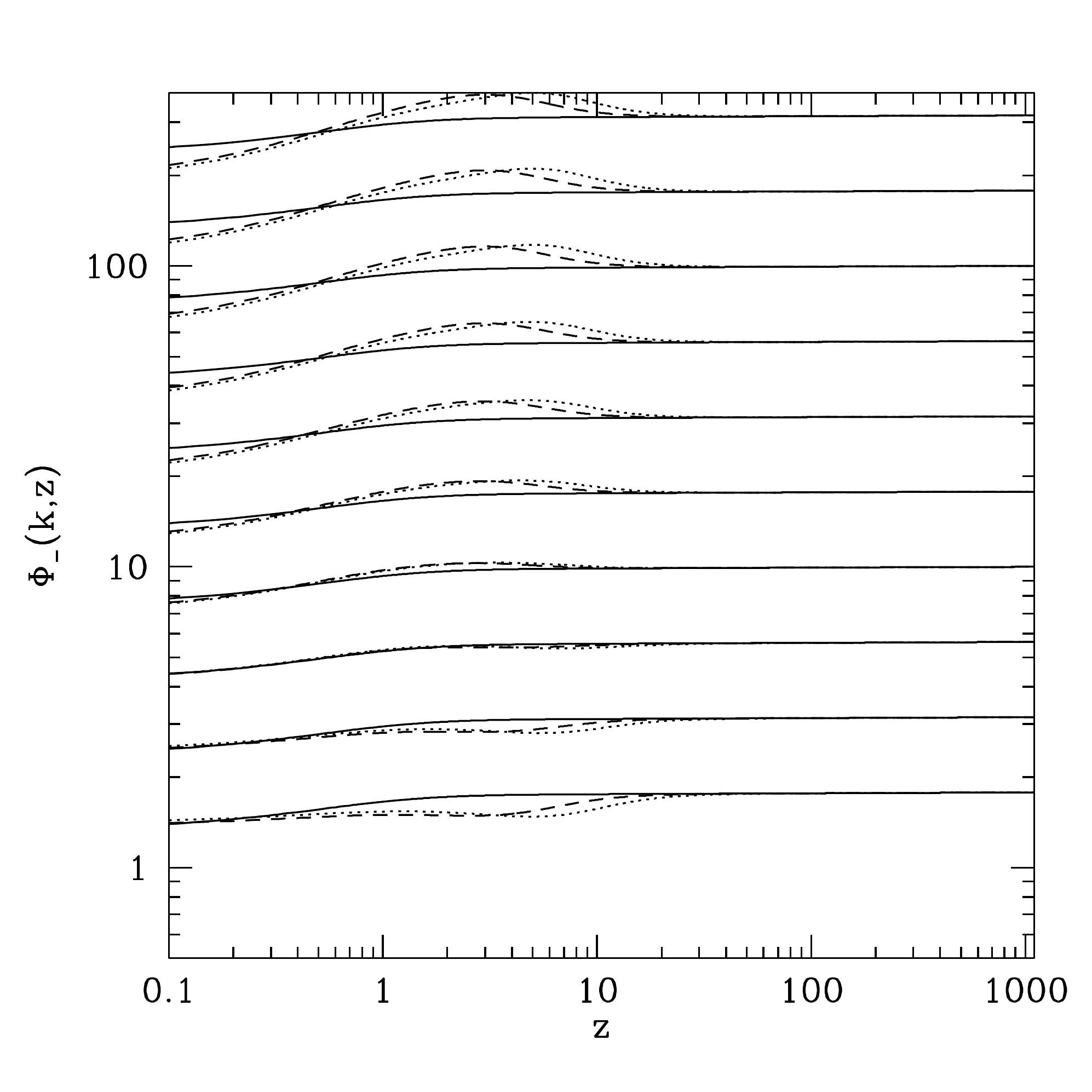}
\includegraphics[width=\columnwidth]{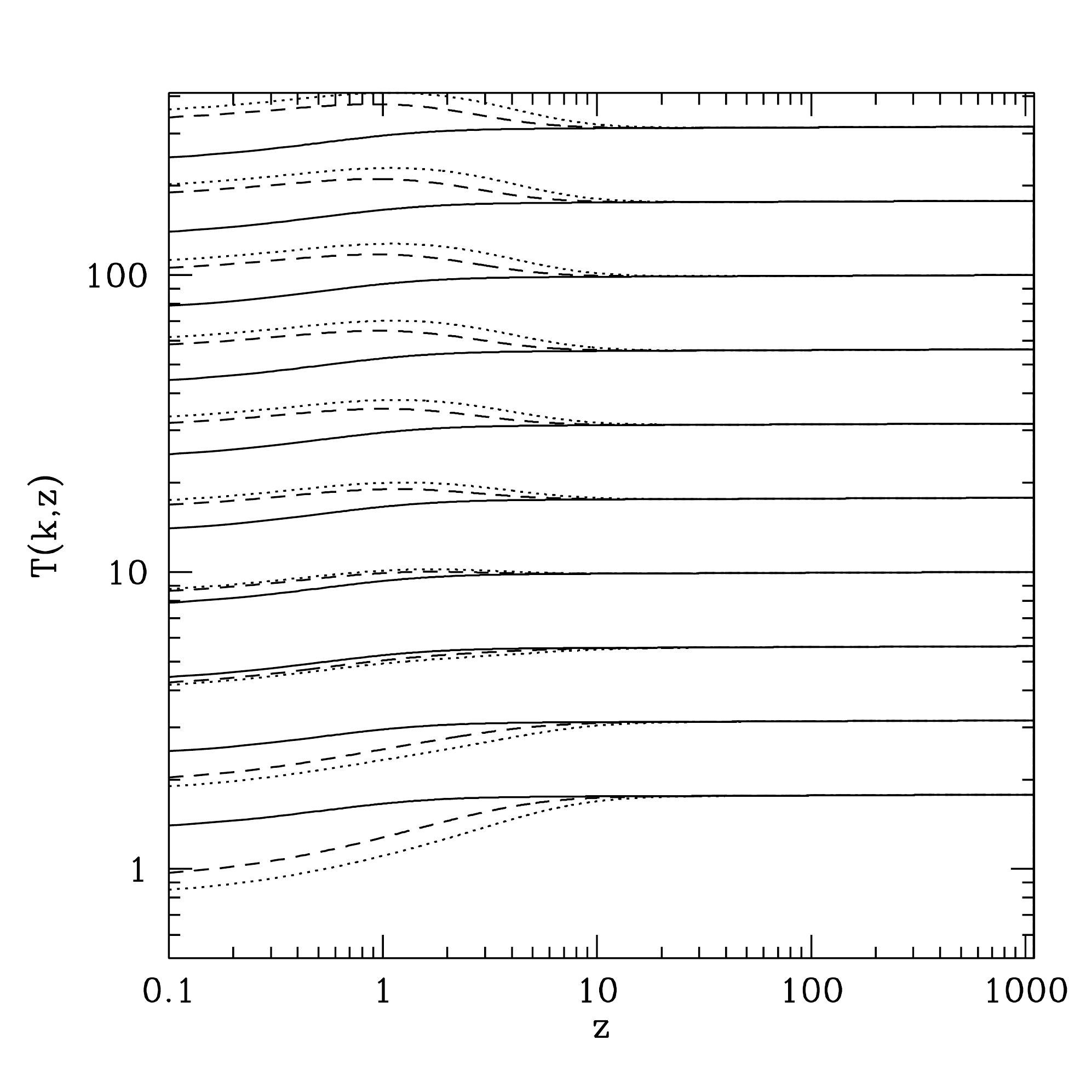}
\caption{Transfer functions for the lensing potential $\Phi_-$ and the comoving density perturbations $\Delta_{\rm m}$, as a function of redshift for $\Lambda$CDM (solid), $r_c = 300$ Mpc (dotted) and $r_c = 600$ Mpc (dashed), all assuming $D=6$. The curves show $k=10^{-4}, 3.6 \times 10^{-4}, 1.3 \times 10^{-3}, 4.6 \times 10^{-3}, 1.7 \times 10^{-2}, 6 \times 10^{-2}, 0.2, 0.8, 3$, and $10~{\rm Mpc}^{-1}$ from bottom to top, and are displaced for clarity. Prior to the domination of $\Lambda$, large scale (super-horizon) modes decay, while small-scale (sub-horizon) modes grow with time. Once $\Lambda$ dominates, all modes start to decay with time.}
\label{transf}
\end{figure*}

Figure~\ref{transf} demonstrates this behavior for $\Phi_-$ and $\Delta_{\rm m}$ with $D=6$ over the range $k=10^{-4}-10$ Mpc$^{-1}$
covering both super-horizon and sub-horizon modes. (What is plotted in the right panel can be thought of as the transfer function for an effective potential related to $\Delta_{\rm m}$ through the standard Poisson equation.)
As expected, prior to the domination of $\Lambda$, large scale (super-horizon) modes decay, while small-scale sub-horizon modes grow with time.

Once $\Lambda$ dominates, however, all modes start to decay. Indeed, since $H' \rightarrow 0$ in this regime, from~(\ref{zetacons}) we find:
\beq
\Phi'_-+\left(1-g\over 1+g\right)\Phi_ -=0 \; \Longrightarrow \; \Phi_- \propto a^{-\frac{1-g}{1+g}}\,.
\eeq
Thus, while $\Phi_-$ decays as $a^{-1}$ at late times in $\Lambda$CDM cosmology, here it decays even faster, as $a^{-D+3}$, since $g=-(D-4)/(D-2)$ on small scales in our model.
In particular, for $D=6$, we have $\Phi_-\sim a^{-3}$ at late times. This behavior can be seen in Fig.~\ref{transf}, where $\Phi_-$ initially grows by up to $\sim 10\%$
as the graviton Compton wavelength enters the horizon, in agreement with~(\ref{Phi109}), but then decays faster than in the $\Lambda$CDM model after the onset of cosmic acceleration. The competition between these two effects is why, as we will see below, gravitational weak lensing is not substantially affected by the graviton Compton wavelength in our model at low redshifts.

\begin{figure}
\includegraphics[width=\columnwidth]{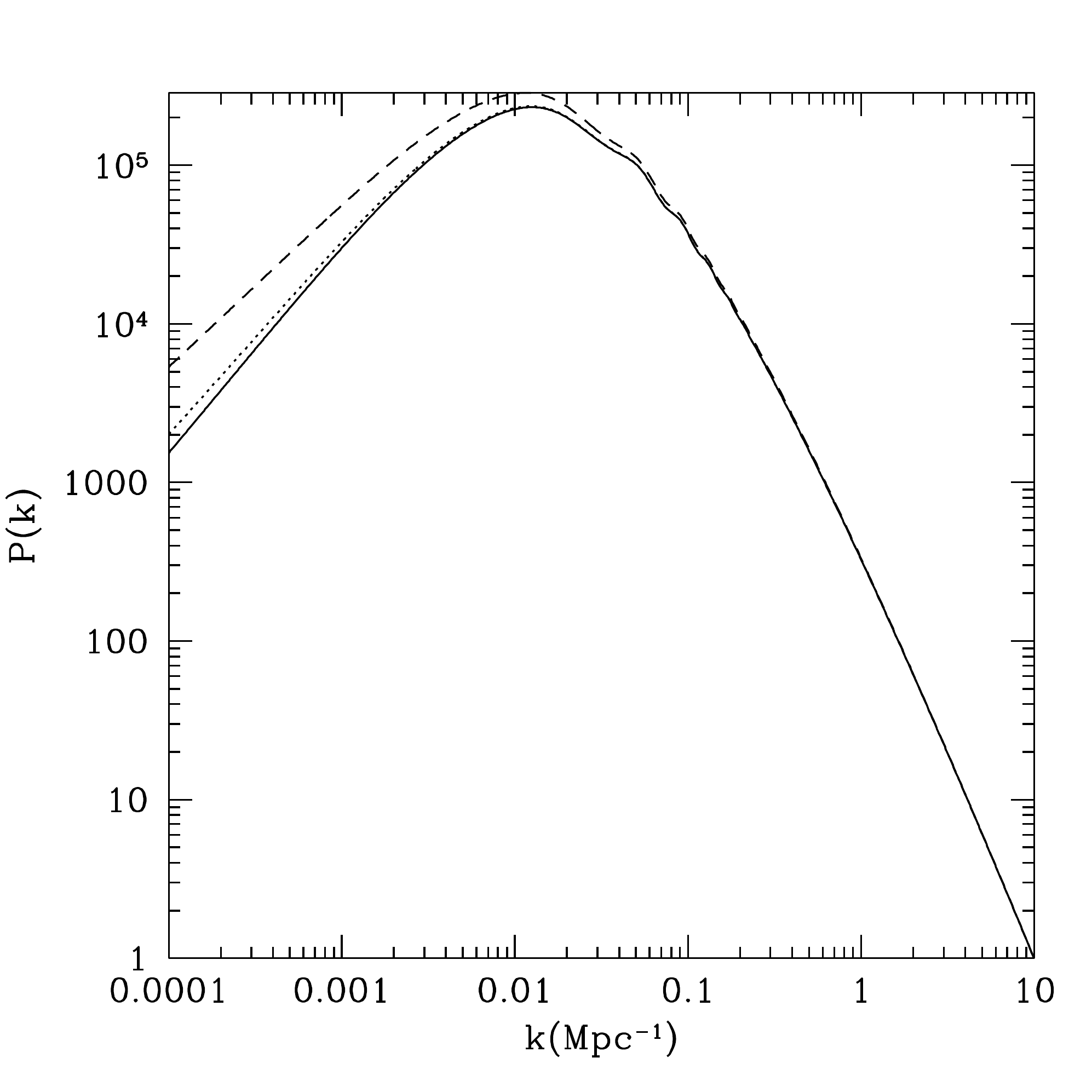}
\caption{The power spectrum of the linear comoving density $\Delta_{\rm m}$ at $z= 0.1$ (solid), $z=1$ (dotted), and $z=5$ (dashed), for $r_c=300$ Mpc. For clarity, the spectra are normalized to unity at $k= 10$ Mpc$^{-1}$. We see that, unlike the $\Lambda$CDM model, the shape of the power spectrum evolves on large scales, which will cause the galaxy bias to be generically scale-dependent.}
\label{power_matter}
\end{figure}

In the following sections, we will discuss the predictions of our model for different measures of structure formation at low redshifts. One notable omission is the clustering of galaxy redshift surveys, such as SDSS and 2dF, which have been instrumental in measuring the shape of the matter power spectrum at low redshifts. These studies rely on a key assumption: that galaxy distribution follows matter distribution on large scales, up to an arbitrary but {\it constant bias} factor. Unfortunately the assumption of constant bias only holds if the shape of the linear matter power spectrum does not change with time, as is the case in the $\Lambda$CDM model. Here, on the other hand, the shape of the matter power spectrum does evolve in time, as shown in Fig.~\ref{power_matter}.
As pointed out in~\cite{Hui:2007zh},  we therefore expect a scale-dependent bias, since the shape of the power spectrum of galaxies may depend on that of matter during the entire history of galaxy formation. We shall postpone a full study of this effect, and only warn that even the large scale bias in modified gravity models cannot be captured by a single number.

\subsection{Weak lensing power spectrum}
\label{WL}

\begin{figure}
\includegraphics[width=\columnwidth]{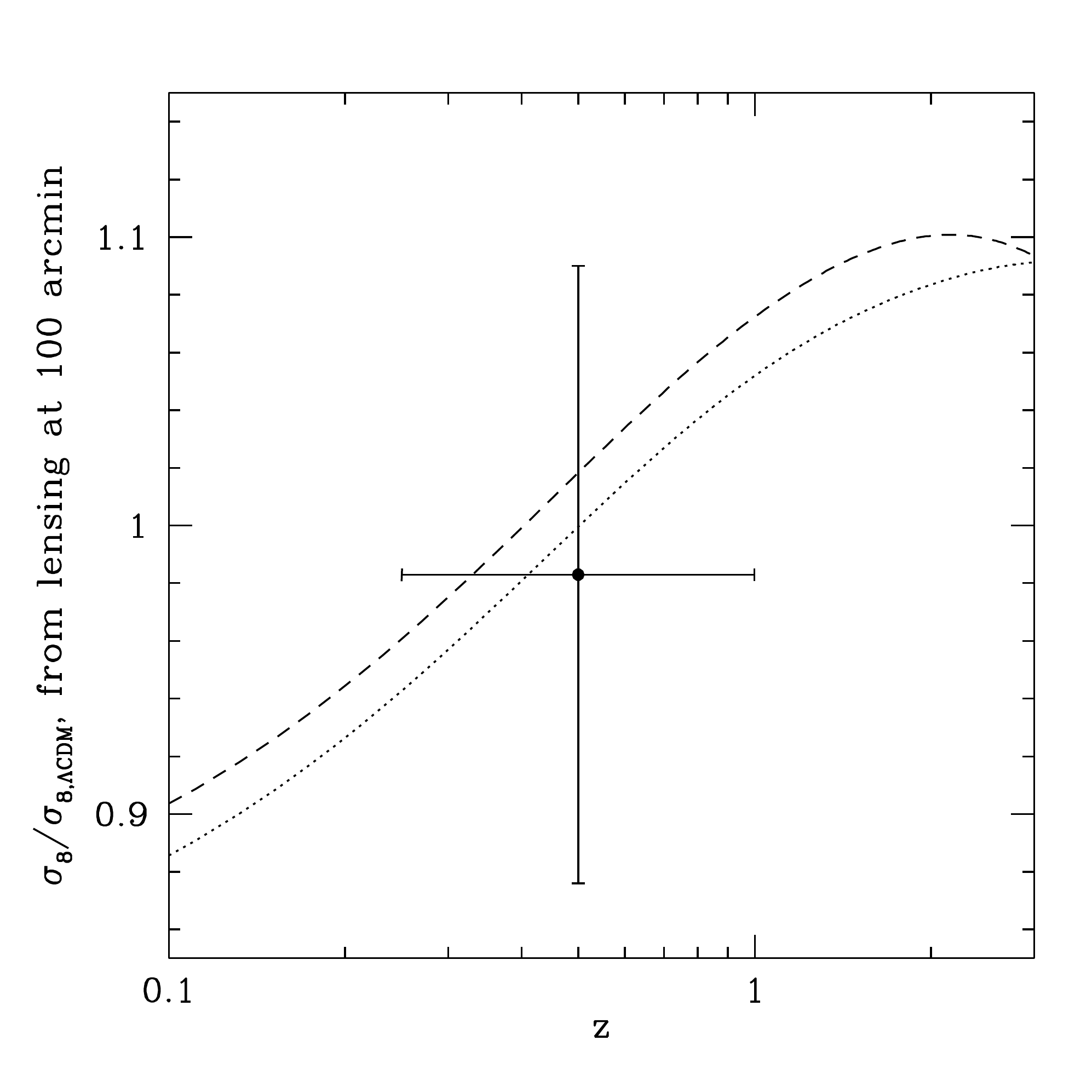}
\caption{The ratio of lensing correlation measured within $100'$ radius for $r_c=300$ Mpc (dotted) and $r_c=600$ Mpc (dashed) with $D=6$, to the $\Lambda$CDM prediction. The data point is from CFHTLS Wide weak lensing measurements~\cite{Fu:2007qq}, with the vertical error bar denoting the 1$\sigma$ range. The horizontal error bar spans the redshift range for the relevant lensing kernel.}
\label{s8_lens}
\end{figure}

Weak lensing is one of the few direct probes of the large-scale gravitational potential with minimal confusion from complicated astrophysical processes. While early weak lensing measurements were plagued by observational systematics, there have been much recent improvement in controlling these errors. Here we will focus on the recent measurement of large-angle weak lensing correlations with Canadian France Hawaii Telescope (CFHT) observations \cite{Fu:2007qq}, which finds:
\beq
\sigma_8 \left(\Omega_{\rm m}\over 0.25\right)^{0.53} =0.837 \pm 0.084
\eeq
on angular scales $85' < \theta < 230'$. Comparing this with the WMAP5 combined constraints with baryonic acoustic oscillations (BAO) in galaxy surveys and type Ia supernovae~\cite{Komatsu}, we see that the lensing amplitude is $0.98 \pm 0.11$ times the CMB-preferred value, {\it i.e.} they are in excellent agreement. Figure~\ref{s8_lens} compares this result with predictions of our massive gravity model. To model the CFHT weak lensing measurements, we plug $\Phi_-$ in our model into the standard Poisson equation, and compute the resulting density fluctuations within a sphere of angular radius $100'$, roughly corresponding to the scale of weak-lensing measurements. As the CFHT lensing sources are broadly distributed in the redshift range $0.5-2$, we assume the range $z= 0.25-1$ for the lensing kernel.

Therefore, due to the competition between massive gravity effects and cosmic acceleration discussed above, we see that current weak lensing measurements cannot distinguish our predictions from that of the $\Lambda$CDM model. However, future  percent-level tomographic measurements of weak lensing power from the Large Synoptic Survey Telescope~\cite{LSST}, Pan-STARRS~\cite{panstarrs}, or NASA's Joint Dark Energy Mission~\cite{JDEM} experiment will be able to test our predictions for the lensing potential.

\subsection{Integrated Sachs-Wolfe Cross Correlation}
\label{ISW}

\begin{figure}
\includegraphics[width=\columnwidth]{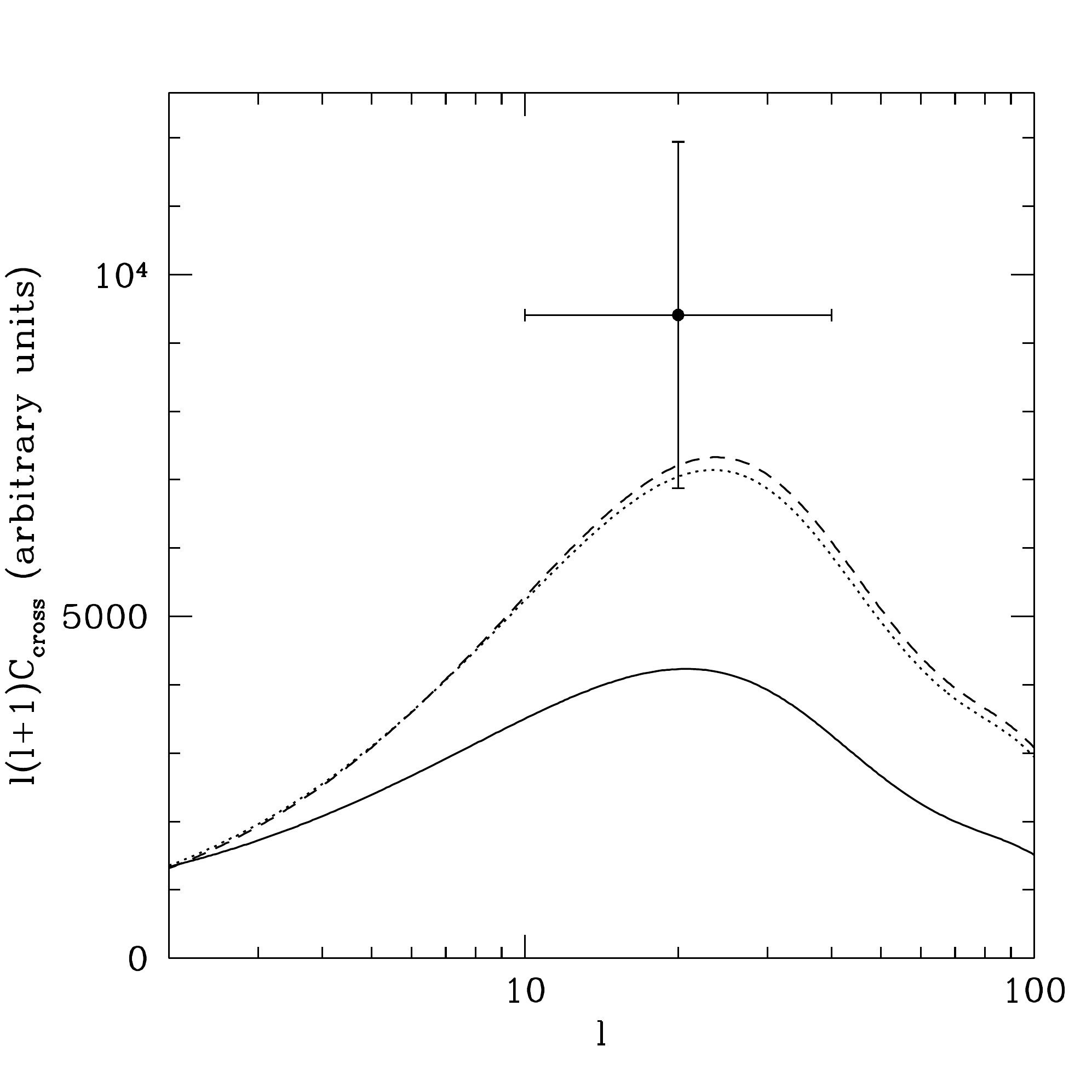}
\caption{Cross-power spectrum of a galaxy survey at $z \simeq 0.5$ with the CMB, for $\Lambda$CDM (solid), $r_c = 300$ Mpc (dotted) and $r_c = 600$ Mpc (dashed), again for $D=6$. The data point is a rough measure of the current observations \cite{seljak} (see the text). The horizontal error bar spans the relevant angular range.}
\label{isw_fig}
\end{figure}

The cross-correlation between the ISW effect in the CMB and the large scale structure at low redshifts probed by galaxy surveys, was first proposed in \cite{Crittenden:1995ak} and later detected by several groups in cross-correlation with WMAP CMB maps. Most recently, \cite{seljak} and \cite{Giannantonio:2008zi} combined different galaxy surveys to obtain an overall detection of the ISW effect in cross-correlation at the $4-4.5\sigma$ level, within the redshift range $z= 0.1-1.5$. However, it was also pointed out in \cite{seljak} that the amplitude of this detection is $2.23 \pm 0.60$ larger than what is predicted by $\Lambda$CDM model (see also \cite{Neyrinck} for a similar result, but at lower significance). Going back to Fig.~\ref{transf}, we note that a faster decay of $\Phi_-$ is indeed expected in our model at low redshifts. In Fig.~\ref{isw_fig}, we compare our prediction for the ISW cross-power spectrum for a galaxy survey at $z \simeq 0.5$, with that of $\Lambda$CDM. The data point, associated with the excess observed amplitude, is centered at $\ell \sim 20$, as most of the $S/N$ in ISW detection in cross-correlation comes for $z\sim 0.5$ and $\ell \sim 10-40$ \cite{Afshordi:2004kz}. Since the ISW effect is proportional to $\dot{\Phi}_-$, as seen in~(\ref{isw_sw}), we have scaled the $\Lambda$CDM prediction for cross-power by the ratio of $\dot{\Phi}_-$ for our model to the standard one, in order to obtain the dotted and dashed curves in Fig.~\ref{isw_fig}. Additional correction due to the changes in the galaxy overdensity (which is correlated with the ISW) is possible, but since the bias is obtained by fitting the observed galaxy clustering in the present model comparisons \cite{seljak,Giannantonio:2008zi}, these corrections are expected to be small.

Interestingly, as can be seen from Fig.~\ref{isw_fig}, the signal is fairly insensitive to the value of the graviton Compton wavelength over the range of interest. This is because, for these values of $r_c$,
 the transition to the modified gravity regime when $H\sim r_c^{-1}$ is already completed by $z=0.5$. Therefore, we see that massive/extra-dimensional
 gravity can explain the observed excess in the ISW effect detection in cross-correlation.

\subsection{Bulk Flows}
\label{BF}

\begin{figure}
\includegraphics[width=\columnwidth]{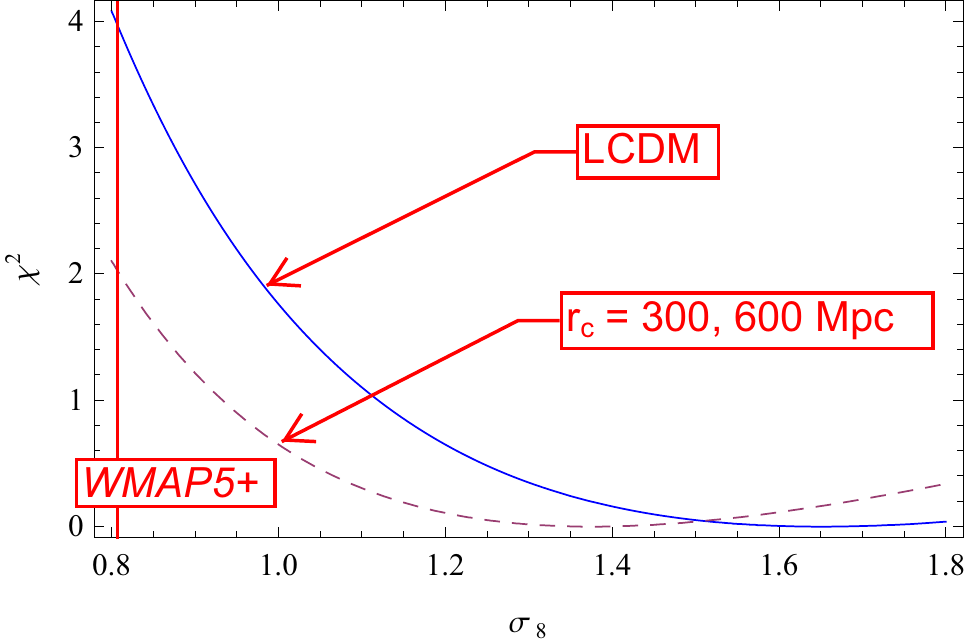}
\caption{The relative $\chi^2$ vs $\sigma_8$ in $\Lambda$CDM and our modified gravity models for producing the observed bulk flows on $\sim 100 $ Mpc scales \cite{hudson}. The vertical line shows the WMAP5+BAO+SN best fit value.}
\label{bulk}
\end{figure}

Another measure of structure growth on large scales are peculiar velocity measurements, which, through the continuity equation, probe $\dot{\Delta}_{\rm m}$ in the linear regime. As discussed in the Introduction, recent measurements of coherent peculiar velocity (or bulk flows) on scales of $\sim 100$~Mpc, from very different methods \cite{hudson,kash}, have led to consistent magnitudes and directions which are nevertheless significantly larger than the $\Lambda$CDM prediction. In particular, \cite{hudson} find a mean velocity of $407 \pm 81$ km/s within a gaussian filter of radius 50~$h^{-1}$Mpc, centered on the Milky Way, while the $\Lambda$CDM WMAP5 best-fit model predicts an r.m.s. value of $190$ km/s. Figure~\ref{bulk} shows how the $\chi^2$ ($= -2$ log of likelihood) for this measurement depends on the $\sigma_8$ value of the $\Lambda$CDM model, keeping all other parameters fixed to their WMAP5 best-fit values. We see that the standard value $\sigma_8 = 0.812 \pm 0.026$ is disfavored at about 2$\sigma$. However, both of our fiducial models with $r_c =300$ and $600$ Mpc increase the r.m.s. velocity on $100$~Mpc scale by $21\%$, which brings down the level of disagreement to $1.4\sigma$, as shown in Fig.~\ref{bulk}.

\subsection{Lyman-$\alpha$ forest}
\label{lalpha}

\begin{figure}
\includegraphics[width=\columnwidth]{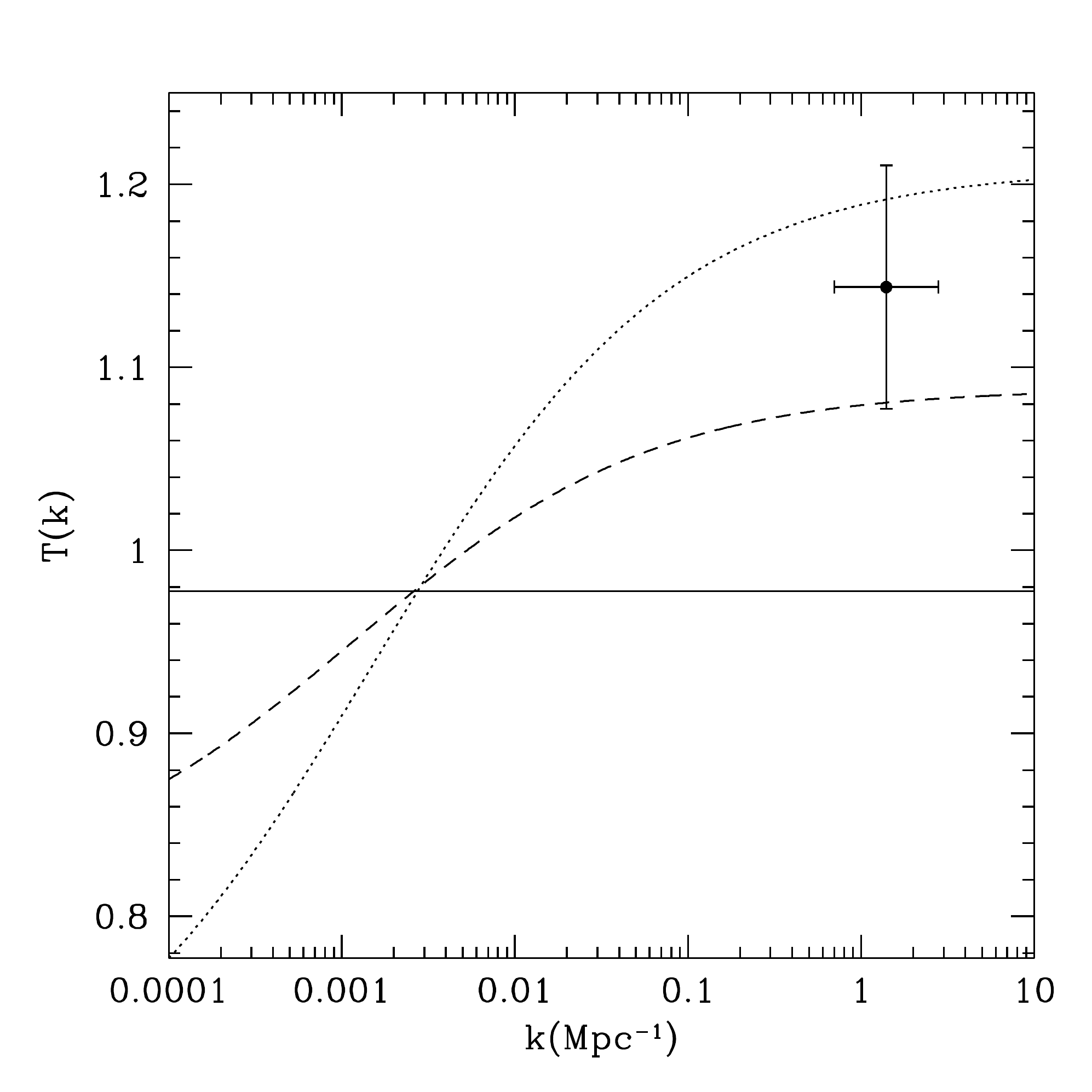}
\caption{The transfer function of the density potential (related to $\Delta_{\rm m}$ through the standard Poisson equation) at $z=3$ and $k=1.4$ Mpc$^{-1}$ for $\Lambda$CDM (solid), $r_c = 300$ Mpc (dotted) and $r_c = 600$ Mpc (dashed). The data point (with 1$\sigma$ error bar) characterizes the excess power observed in Lyman-$\alpha$ forest observations \cite{Seljak:2006bg}. The horizontal error bar approximates the range of scales probed by these observations.}
\label{Ly-alpha}
\end{figure}

The absorption spectrum of the Lyman-$\alpha$ transition in the intergalactic neutral hydrogen on the line of sight to high redshift quasars, commonly known as the Lyman-$\alpha$ forest, has so far provided us with our most stringent constraints on the small-scale power spectrum of matter perturbations at $z=2-3$. The interpretation of these observations is complicated by mild non-linear evolution, as well as baryonic physics and other features in the quasar spectra, and thus large suits of numerical simulations have been used to calibrate the relation of the observed flux spectrum to the underlying linear matter perturbations \cite{McDonald:2004eu, Viel:2005ha}. In this section, we compare the predictions of our model to the constraints obtained on the linear theory power spectrum at $z=3$ and $k \simeq 1.4$ Mpc$^{-1}$ in \cite{McDonald:2004xn}. Seljak, Slosar \& McDonald~\cite{Seljak:2006bg} used these constraints to conclude that the predicted linear power spectrum from the WMAP3 best-fit model was too low to explain the observed spectrum of Lyman-$\alpha$ forest at the $\sim 2.4\sigma$ level. They attributed this excess to the presence of extra neutrinos (or other relativistic components) during radiation-matter equality. Their best fit model with extra neutrinos has roughly $35 \pm 14\%$ more power than the standard $\Lambda$CDM WMAP3 best fit model with 3 neutrinos. While a consistent interpretation of this discrepancy in the context of our model requires a similar suit of simulations with a full treatment of non-linear modified gravity, as a crude first step Fig.~\ref{Ly-alpha} compares this excess with our model prediction for the small-scale growth in linear density $\Delta_{\rm m}$. Interestingly, either fiducial values of the graviton Compton wavelength in our model ($r_c =300$ or 600~Mpc) can explain the observed excess.

\subsection{Galaxy clusters and CBI excess}
\label{CBISZ}

\begin{figure}
\includegraphics[width=\columnwidth]{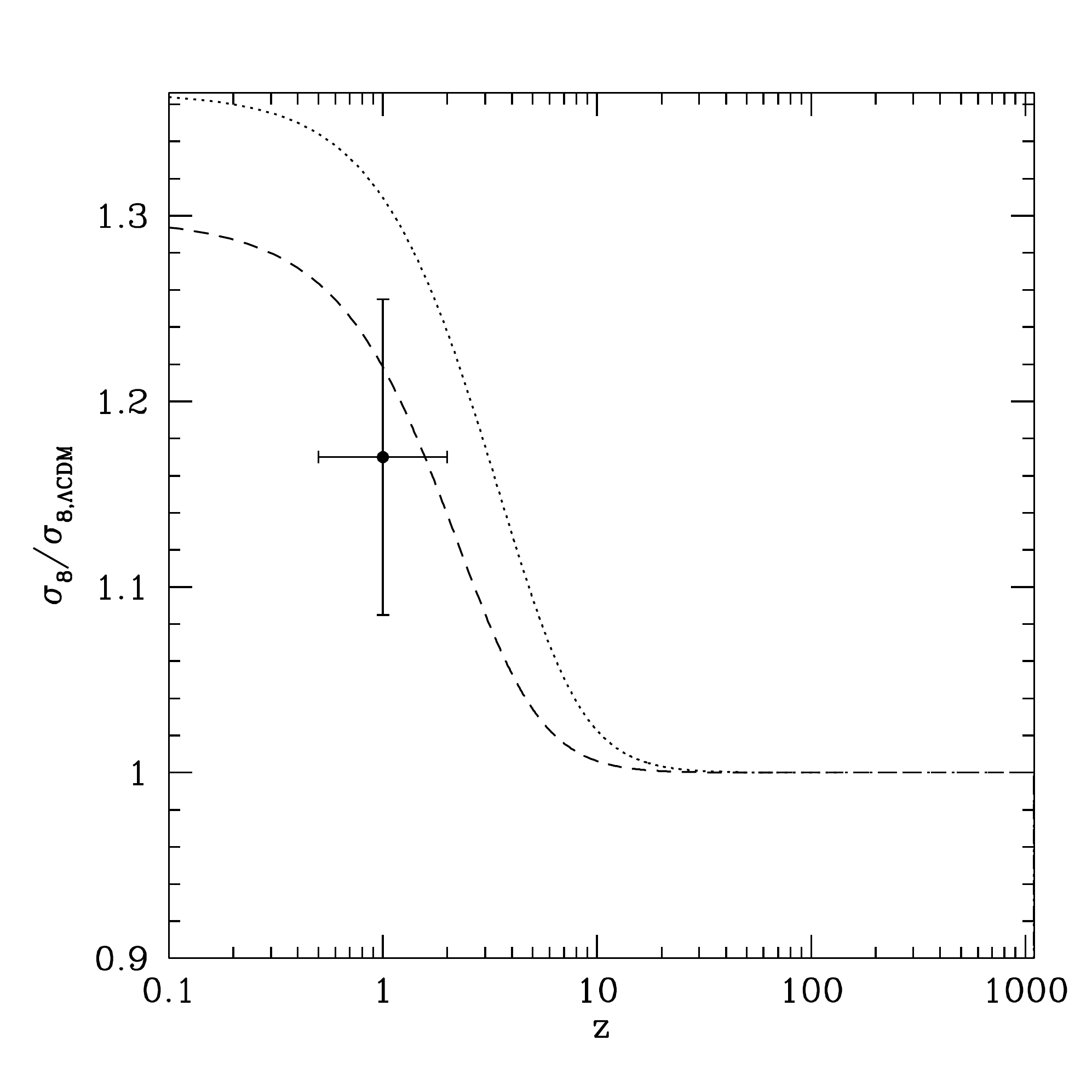}
\caption{The ratio of $\sigma_8$ in our modified gravity models to that of $\Lambda$CDM, as a function of redshift. The dotted and dashed curves correspond to $r_c = 300$ and $600$ Mpc respectively. The data point shows the observed excess associated with the CBI excess \cite{Reichardt:2008ay}.}
\label{s8_dgp}
\end{figure}

Finally, we will address another long-standing puzzle in CMB observations, known as the CBI excess. The high resolution observations of the Cosmic Background Imager (CBI) have long indicated an excess in the power spectrum of CMB anisotropies at $\ell \sim 3000$, far above the expected level of primary anisotropies \cite{Readhead:2004gy}. Consistent results have been obtained by BIMA \cite{Dawson:2006qd} and ACBAR \cite{Reichardt:2008ay} experiments, albeit at lower significance levels. While this excess has a consistent spectrum with that of the thermal SZ effect --- the dominant source of CMB anisotropy at $\ell \sim 3000$ ---, the observed level is significantly higher than what is expected in the WMAP5 best-fit $\Lambda$CDM cosmology. The amplitude of the SZ power spectrum is a direct measure of the abundance of massive galaxy clusters at $z \sim 0.5-2$, and scales as $\sigma_8^7$ \cite{Komatsu:2002wc}. The current excess in the SZ power can be interpreted as a $17\%$ larger value of $\sigma_8$ at $2-3\sigma$ level \cite{Reichardt:2008ay}. As $\sigma_8$ is the amplitude of the linear matter density fluctuations close to the scale of cluster formation, we expect to capture most of the physics relevant for cluster abundances by tracking the evolution of $\sigma_8(z)$ in our model. Figure~\ref{s8_dgp} shows our prediction for the excess of $\sigma_8(z)$ compared to the $\Lambda$CDM model, which could easily explain the CBI excess.

We should note that the observed deficit of the SZ decrement in WMAP clusters \cite{Afshordi:2006pc} --- indicating a smaller SZ flux per cluster --- may point to an even larger excess required to explain the CBI excess. Future observations of SZ power spectrum as well as SZ clusters by the Atacama Cosmology Telescope (ACT) and the South Pole Telescope (SPT) should be able to decisively confirm or rule out this excess.

A seemingly contradictory result comes from the mass function of galaxy clusters at lower redshifts, inferred from X-ray observations. Most recently, Vikhlinin et al. \cite{vikhlinin} showed that Chandra observations of a sample of galaxy clusters detected in X-rays by ROSAT are remarkably consistent with the predictions for mass function from $\Lambda$CDM, and in particular $\sigma_8 = 0.77 \pm 0.02$ (including systematic errors, but using geometrical constraints on $\Omega_m$). If correct, this would be inconsistent with the common interpretation of the CBI excess as a large $\sigma_8$ \cite{Komatsu:2002wc}, and may appear to rule out our preferred range of $r_c=300-600$~Mpc (Fig.~\ref{s8_dgp}). However, the relationship between the ROSAT X-ray luminosities and the cluster masses has a large scatter, which is not constrained very well empirically (see Fig. 13 in \cite{vikhlinin_analysis}). In particular, an extended low-luminosity tail in the distribution of luminosities for a given mass would mean that a significant number of massive clusters may go undetected, leading to a possible underestimate of $\sigma_8$.

We also should point out that our analysis so far is only strictly valid in the linear regime and the formation of non-linear collapsed objects (such as galaxy clusters) is not modeled in our framework. In other words, the relationship between the number of clusters and the linear power spectrum, used in \cite{vikhlinin}, is only valid for $\Lambda$CDM cosmology, and could be significantly different in modified gravity models. Indeed, as we discussed in \ref{moon}, this is expected due to the Vainshtein effect at large densities, and is required in order to satisfy the precision tests of gravity at solar system and galactic scales. Therefore, a good understanding of non-linear collapse process and Vainshetin effect in our model is necessary to make reliable predictions for the cluster mass function (and CBI excess). 

\section{Conclusions}
\label{conc}

Arguably the most pressing question in cosmology is whether the observed cosmic acceleration
is due to a breakdown of Einstein gravity or to conventional vacuum energy. Fortunately,
the $\Lambda$CDM model is highly predictive --- a single dark energy parameter $\Lambda$ suffices
to predict the entire expansion and growth histories. Near future probes of the large scale structure
will subject this standard model to increasingly stringent tests of its predictions.

In this paper we have presented a case that current observations may already be pointing towards new gravitational physics
on cosmological scales. The excess of power seen in the Lyman-$\alpha$ forest and small-scale CMB experiments, the larger
bulk flows seen both in peculiar velocity surveys and in kinetic SZ, and the higher ISW cross-correlation all present potential challenges for
$\Lambda$CDM. And although their individual statistical significance is only at the $\sim 2\sigma$
level, it is certainly intriguing that these observations suggest a common explanation: that structure is more evolved than expected
from $\Lambda$CDM. On the other hand, weak lensing measurements give a value for $\sigma_8$ consistent with the CMB.

We have argued that these observations find a natural explanation in models where the graviton is
a resonance with a tiny width $r_c\sim 300-600$~Mpc. The longitudinal mode $\pi$ of the graviton can be thought
of as mediating an extra scalar force which, thanks to the Vainshtein effect, only kicks in at late times
and on sufficiently large scales. This speeds up structure formation, thereby accounting
for the above anomalies. Meanwhile, since photons do not couple directly to $\pi$, weak lensing is much less affected, which
explains its consistency with the CMB.

Our approach has been motivated by higher-dimensional generalizations of the DGP scenario, such as Cascading Gravity models.
A crucial advantage of considering more extra dimensions is that the modifications to the Friedmann equation become indistinguishable
from a cosmological constant. This makes it possible to consider much smaller values of the graviton Compton wavelength than
usually allowed by constraints on the expansion history. And this in turn amplifies the impact of the longitudinal mode on structure formation.

The obvious danger with enhancing structure growth is that one risks boosting the ISW contribution to the CMB low-$\ell$ power.
In fact, observations show the opposite trend, starting with the well-known low quadrupole. A more statistically significant anomaly is the
peculiar fact that the angular correlation function vanishes on $\gsim 60\,^{\circ}$ scales, as seen both in COBE and WMAP.

We have shown that our modified gravity models can account for these large scale CMB anomalies. With certain assumptions about the
super-horizon behavior of $\pi$, we have presented a novel mechanism in which a late-time ISW contribution cancels against the
primordial Sachs-Wolfe amplitude. This ISW contribution is different from the usual one due to $\Lambda$, but arises because
the dynamics of $\pi$ undergo a transition when $H\sim r_c^{-1}$. Our mechanism relies on conservation of $\zeta$ and is analogous
to the jump in the Newtonian potential incurred during the radiation- to matter-dominated era in standard cosmology.

Many issues remain to be addressed:

\begin{itemize}

\item To compare our predictions with galaxy surveys, such as SDSS, we need a handle on the halo bias
in our model. As argued in~\cite{Hui:2007zh}, the bias is generically expected to be scale dependent in modified gravity theories.
This issue is currently under study using N-body simulations~\cite{mark}.

\item A more thorough comparison with data requires a full likelihood analysis, including the PPF parameters used here.

\item The effects of modified gravity is suppressed by the Vainshtein effect at large densities. How does this affect the non-linear collapse process, and the halo mass function in our model? 

\item While the phenomenological approach adopted here constitutes an important and instructive first step,
it is imperative to justify the choice of PPF input functions through full-fledged brane-world calculations, for instance
in Cascading Gravity models. In particular, the suppression of the large-angle CMB power relies
on certain assumptions about the behavior of perturbations on super-Hubble scales, as mentioned above, and it
will be essential to check whether this behavior can be explicitly realized in higher-dimensional constructions.
If not, are there other consistent theories of modified gravity that can reproduce the phenomenology presented here?

\end{itemize}

\section*{Acknowledgements}
The authors would like to thank Claudia de Rham, Amir Hajian, Mike Hudson, Pat McDonald, P.J.E. Peebles, Roman Scoccimarro, David Spergel, Glenn Starkman, Andrew Tolley, Alexey Vikhlinin, and Mark Wyman for helpful comments and discussions. We thank Wenjuan Fang for making the implementation of the PPF formalism in CAMB publicly available.
This work was supported by the Perimeter Institute for Theoretical Physics.  Research at Perimeter Institute is supported by the Government of Canada through Industry Canada and by the Province of Ontario through the Ministry of Research \& Innovation.

\appendix
\section{Review of PPF Formalism}\label{ppf}

In this Appendix we review the essential ingredients of the PPF formalism, considering the more general case where $\zeta$ is not necessarily conserved
on sub-Hubble scales. Following~\cite{ppf} we study the matter perturbations in comoving gauge. In this gauge, the perturbed stress tensor for dark matter
has non-zero components $T^0_{\; 0} = -\rho(1+\Delta_{\rm m})$ and $T^0_{\; i}  = -\rho\partial_i v_{\rm m}/ka$, where $\Delta_{\rm m}$ and $v_{\rm m}$
are by definition the energy density and velocity perturbations, respectively.

The matter stress-energy tensor is assumed to be covariantly conserved, as is the case in the DGP model and its higher-codimension extensions.
Conservation of energy and momentum then implies
\bea
\nonumber
\Delta_{\rm m}' + \frac{k}{aH}v_{\rm m} &=&-3\zeta'\;; \\
v_{\rm m}' + v_{\rm m} &=& (g-1)\frac{k}{aH}\Phi_-\,,
\label{Emomap}
\eea
with $'\equiv {\rm d}/{\rm d}\ln a$. Here $\zeta$ denotes the curvature perturbation, related to the other variables by
\beq
\zeta = \Phi - \frac{aH}{k}v_{\rm m}\,.
\label{zetadef}
\eeq

\subsection{Super-horizon evolution}

Assuming adiabatic initial conditions, the curvature perturbation, $\zeta$, must be conserved for long wavelengths.
As shown by Bertschinger~\cite{ed}, this yields a consistency relation valid on super-Hubble scales:
\bea
\nonumber
& & \zeta = {\rm const.} = \frac{H}{H'}\left[(g-1)\Phi_- -g'\Phi_- - (g+1)\Phi_-'\right] \\
& & \;\;\;\;\;\;\;\;\;\;\;\;\;\;\;\;\;\;\;\;\;\;\;\;\;\;\;+ (g+1)\Phi_-\,.
\label{zetaap}
\eea
It follows that the evolution of super-horizon modes is uniquely determined by specifying $g$ and the expansion history.

\subsection{Sub-horizon evolution}

On sub-Hubble scales, $k\gg aH$, we enter the Newtonian regime, in which metric and matter perturbations are related
by a modified Poisson equation
\beq
k^2\left[1+f(a,k)\right]\Phi_- = 4\pi Ga^2\Delta_{\rm m}\,.
\label{poissonmain}
\eeq
Matching the above form factor with our modified propagator~(\ref{prop}) fixes $f(a,k)$:
\beq
f(a,k) = \left(\frac{a}{kr_c}\right)^{2(1-\alpha)}\,.
\label{fdef}
\eeq
Note that $f$ has been neglected in most earlier studies of DGP cosmological perturbations~\cite{lue,koyama,song}, except in~\cite{amin}, presumably because $r_c$ was assumed to be of order $H_0^{-1}$, and therefore all modes of interest satisfied $k\gg ar_c^{-1}$. In this work, however, we have considered much smaller values for $r_c$, in which case $f(a,k)$ must be carefully taken into account.

On the matter side, the energy-momentum conservation equations~(\ref{Emomap}) together yield an evolution equation for $\Delta_{\rm m}$,
valid in the Newtonian regime:
\beq
\Delta_{\rm m}'' + \left(2+\frac{H'}{H}\right)\Delta_{\rm m}'=\frac{k^2}{a^2H^2}(1-g)\Phi_-\,.
\label{deleomap}
\eeq
This can then be combined with~(\ref{poissonmain}) to derive a second-order equation for $\Phi_-$.

\subsection{Interpolating between sub- and super-horizon regimes}

In general, of course, the resulting sub-horizon evolution equation for $\Phi_-$ will be inconsistent with conservation of $\zeta$ when applied in
the long-wavelength limit. One must therefore interpolate between the sub- and super-horizon behaviors derived above.
To do so, following~\cite{ppf} we introduce a new function $\Gamma(a,k)$ through a modified Poisson equation, valid {\it on all scales}:
\beq
k^2(\Phi_- + \Gamma) = 4\pi Ga^2\rho\Delta_{\rm m}\,.
\label{Gam}
\eeq

At long wavelengths, $k\ll aH$, we require that $\Gamma$ enforces the conservation of $\zeta$.
By taking the time-derivative of~(\ref{Gam}) and combining with the conservation equations~(\ref{Emomap}), we obtain
\beq
\Gamma'+\Gamma = -\Phi_-' - \Phi_- - \frac{3}{2}\Omega_{\rm m}(1+f_\zeta)\frac{aH}{k}v_{\rm m}\,,
\eeq
where $f_\zeta$ specifies the leading behavior of $\zeta'$ as $k\rightarrow 0$:
\beq
\zeta' \rightarrow \frac{1}{3}f_\zeta\frac{k}{aH}v_{\rm m}\,.
\eeq
Then, taking the time-derivative of~(\ref{zetadef}), we can substitute for $\Phi_-'$ to obtain an equation {\it valid
for $k\ll aH$}:
\bea
\nonumber
\Gamma' +\Gamma &=& \left(\frac{g'-2g}{1+g}\right)\Phi_-\\
 &-& \frac{aH}{k}v_{\rm m}\left(\frac{H'}{(1+g)H}  + \frac{3}{2}\Omega_{\rm m}(1+f_\zeta)\right).
\label{Gamlong}
\eea

Meanwhile, it is clear from~(\ref{poissonmain}) that $\Gamma\rightarrow f(a,k)\Phi_-$ at short wavelengths ($k\gg aH$).
To interpolate between this condition and the long-wavelength behavior described by~(\ref{Gamlong}), we can
assume that $\Gamma$ satisfies
\bea
\nonumber
&&\left(1+ c_\Gamma^4\frac{k^4}{a^4H^4}\right)\left\{\Gamma' +\Gamma + c_\Gamma^4\frac{k^4}{a^4H^4}\left(\Gamma-f(a,k)\Phi_-\right)\right\} \\
&&\;\;\;\;\;\;\;\;\; = - \frac{aH}{k}v_{\rm m}\left(\frac{H'}{(1+g)H}  + \frac{3}{2}\Omega_{\rm m}(1+f_\zeta)\right)\,.
\label{Gamfull}
\eea
Note that this is very similar to the corresponding equation in~\cite{ppf}, with the only difference that we are using $(c_\Gamma k/aH)^4$
as opposed to $(c_\Gamma k/aH)^2$. A higher power of $k$ is essential here to win over the potential $1/k^2$ dependence of $f(a,k)$, achieved for $\alpha=0$, and thereby
ensuring the correct infrared behavior ($k\rightarrow 0$).

Moreover, we can combine the conservation equations~(\ref{Emomap}) and the Poisson equation~(\ref{Gam}) to obtain
\bwt
\beq
\left(\Phi_-' + \Phi_-\right)\left(\frac{k^2}{a^2H^2} + \frac{9\Omega_{\rm m}}{2}(1+g)\right) = \frac{9\Omega_{\rm m}}{2}(2g-g')\Phi_- -  \frac{k^2}{a^2H^2}(\Gamma'+\Gamma)
+ \frac{3\Omega_{\rm m}}{2}\left(3\frac{H'}{H}-\frac{k^2}{a^2H^2}\right)\frac{aH}{k}v_{\rm m}\,.
\label{Phifull}
\eeq
\ewt
Equations~(\ref{Gamfull}) and~(\ref{Phifull}), together with momentum conservation~(\ref{Emomap}), form a coupled system of first-order equations for
$\Phi_-$, $\Gamma$ and $v_{\rm m}$.

To summarize, the PPF formalism requires specifying an expansion history $H(a)$, a difference in metric potentials $g(a,k)$, and a short-wavelength graviton form factor
$f(a,k)$. Moreover, the transition between sub- and super-Hubble regimes is specified by $c_\Gamma$, and the leading departure from conservation of $\zeta$ is
parametrized by $f_\zeta$. In the particular case $c_\Gamma = f_\zeta = 0$, which was considered in the main text, $\zeta$ is exactly conserved on all scales.
Then, the evolution of $\Phi_-$ is uniquely determined by~(\ref{zetaap}). Meanwhile,~(\ref{deleomap}), which holds exactly in that case, describes the evolution
for $\Delta_{\rm m}$.

\section{Brans-Dicke}\label{Brans-Dicke}

To reassure ourselves that it is perfectly sensible to choose $g>0$ on super-horizon scales, as assumed in~(\ref{g<}), here we show that all stable Brans-Dicke theories have this property. Setting $M_{\rm Pl} = (8\pi G)^{-1/2}=1$, the Brans-Dicke action in Jordan frame takes the form
\beq
S = \int {\rm d}^4x\sqrt{-g}\left(\frac{\sigma R}{2} - \frac{\omega_{\rm BD}}{2\sigma}(\partial\sigma)^2+ {\cal L}_{\rm m}[g] \right) \,,
\eeq
where $\omega_{\rm BD}$ is the Brans-Dicke parameter. In this frame the matter Lagrangian, ${\cal L}_{\rm m}$, is independent of $\sigma$.

For the study of perturbations, however, it is convenient to work instead in Einstein frame, obtained through the conformal transformation:
\beq
g_{{\rm E}\;\mu\nu} = \sigma  g_{\mu\nu} = e^{2\beta\phi} g_{\mu\nu}\,,
\label{conf}
\eeq
where in the last step we have introduced $\beta^2 = 1/(6+4\omega_{\rm BD})$. Moreover, the field redefinition $\sigma = e^{2\beta\phi}$
is such that  $\phi$ is canonically normalized in Einstein frame:
\beq
S =  \int {\rm d}^4x\sqrt{-g_{\rm E}}\left(\frac{R_{\rm E}}{2}- \frac{1}{2}(\partial\phi)^2+ {\cal L}_{\rm m}[g_{\rm E}e^{2\beta\phi}] \right) \,.
\eeq

In the absence of anisotropic stress in the matter sector, one can set $\Phi_E = -\Psi_E$ in Einstein frame. From the conformal transformation~(\ref{conf})
we can therefore read off in Jordan frame, to linear order,
\beq
g = \frac{\Phi+\Psi}{\Phi-\Psi} = \beta\frac{\delta\phi}{\Psi_{\rm E}}\,.
\label{gBD}
\eeq

With the matter described by dust, it is straightforward to show that the cosmological equations of motion allow for a scaling solution
\beq
a_{\rm E}\sim t_{\rm E}^{\;\frac{2}{3+2\beta^2}}\;;\qquad \phi\sim \frac{4\beta}{3+2\beta^2}\log t_{\rm E}\,,
\label{Esolns}
\eeq
where the subscripts make explicit that we are working in Einstein frame. Assuming adiabatic initial conditions, conservation of $\zeta$ on super-horizon scales implies
the following growing-mode solutions
\bea
\nonumber
\Psi_{\rm E} &=& \zeta\left(1-\frac{H_{\rm E}}{a_{\rm E}}\int {\rm d}t_{\rm E} a_{\rm E}\right) = \zeta\cdot\frac{2\beta^2+3}{2\beta^2+5}\;; \\
\delta\phi &=& \zeta \frac{\dot{\phi}}{a_{\rm E}}\int {\rm d}t_{\rm E} a_{\rm E} = \zeta\cdot \frac{4\beta^2}{2\beta^2+5}\,.
\eea
Substituting into~(\ref{gBD}), we obtain
\beq
g\vert_{k\ll aH} = \frac{4\beta^2}{2\beta^2 + 3} = \frac{1}{3\omega_{\rm BD} + 5}\,.
\eeq
Absence of ghosts requires that $3+ 2\omega_{\rm BD} > 0$, and thus $g\vert_{k\ll aH} > 0$ for all stable Brans-Dicke theories, as we were hoping to show.


\begin{thebibliography}{99}

\bibitem{sn1a}
 A.~G.~Riess {\it et al.}  [Supernova Search Team Collaboration],
  ``Observational Evidence from Supernovae for an Accelerating Universe and a
  Cosmological Constant,''
  Astron.\ J.\  {\bf 116}, 1009 (1998);
    S.~Perlmutter {\it et al.}  [Supernova Cosmology Project Collaboration],
  ``Measurements of Omega and Lambda from 42 High-Redshift Supernovae,''
  Astrophys.\ J.\  {\bf 517}, 565 (1999).

\bibitem{seljak}
 S.~Ho, C.~Hirata, N.~Padmanabhan, U.~Seljak and N.~Bahcall,
  ``Correlation of CMB with large-scale structure: I. ISW Tomography and
  Cosmological Implications,''
  Phys.\ Rev.\  D {\bf 78}, 043519 (2008).

 \bibitem{Giannantonio:2008zi}
  T.~Giannantonio, R.~Scranton, R.~G.~Crittenden, R.~C.~Nichol, S.~P.~Boughn, A.~D.~Myers and G.~T.~Richards,
  ``Combined analysis of the integrated Sachs-Wolfe effect and cosmological
  implications,''
  Phys.\ Rev.\  D {\bf 77}, 123520 (2008)
  [arXiv:0801.4380 [astro-ph]].

\bibitem{hudson}
  R.~Watkins, H.~A.~Feldman and M.~J.~Hudson,
  ``Consistently Large Cosmic Flows on Scales of 100 Mpc/h: a Challenge for the
  Standard LCDM Cosmology,''
  arXiv:0809.4041 [astro-ph].

\bibitem{kash}
  A.~Kashlinsky, F.~Atrio-Barandela, D.~Kocevski and H.~Ebeling,
  ``A measurement of large-scale peculiar velocities of clusters of galaxies:
  technical details,''
  arXiv:0809.3733 [astro-ph]; arXiv:0809.3734 [astro-ph].

\bibitem{Readhead:2004gy}
  A.~C.~S.~Readhead {\it et al.},
  ``Extended Mosaic Observations with the Cosmic Background Imager,''
  Astrophys.\ J.\  {\bf 609}, 498 (2004)
  [arXiv:astro-ph/0402359].

\bibitem{Reichardt:2008ay}
  C.~L.~Reichardt {\it et al.},
  ``High resolution CMB power spectrum from the complete ACBAR data set,''
  arXiv:0801.1491 [astro-ph].

\bibitem{McDonald:2004eu}
  P.~McDonald {\it et al.}  [SDSS Collaboration],
  ``The Lyman-alpha Forest Power Spectrum from the Sloan Digital Sky Survey,''
  Astrophys.\ J.\ Suppl.\  {\bf 163}, 80 (2006)
  [arXiv:astro-ph/0405013].

\bibitem{McDonald:2004xn}
  P.~McDonald {\it et al.}  [SDSS Collaboration],
  ``The Linear Theory Power Spectrum from the Lyman-alpha Forest in the Sloan
  Digital Sky Survey,''
  Astrophys.\ J.\  {\bf 635}, 761 (2005)
  [arXiv:astro-ph/0407377].

\bibitem{Seljak:2006bg}
  U.~Seljak, A.~Slosar and P.~McDonald,
  ``Cosmological parameters from combining the Lyman-alpha forest with CMB,
  galaxy clustering and SN constraints,''
  JCAP {\bf 0610}, 014 (2006)
  [arXiv:astro-ph/0604335].

\bibitem{cobe}
 G.~Hinshaw {\it et al.},
  ``2-Point Correlations in the COBE DMR 4-Year Anisotropy Maps,''
  arXiv:astro-ph/9601061.

\bibitem{wmap1}
    D.~N.~Spergel {\it et al.}  [WMAP Collaboration],
  ``First Year Wilkinson Microwave Anisotropy Probe (WMAP) Observations:
  Determination of Cosmological Parameters,''
  Astrophys.\ J.\ Suppl.\  {\bf 148}, 175 (2003)
  [arXiv:astro-ph/0302209].

\bibitem{huterer}
   C.~Copi, D.~Huterer, D.~Schwarz and G.~Starkman,
  ``The Uncorrelated Universe: Statistical Anisotropy and the Vanishing Angular
  Correlation Function in WMAP Years 1-3,''
  Phys.\ Rev.\  D {\bf 75}, 023507 (2007)
  [arXiv:astro-ph/0605135].

\bibitem{others}
 A.~Hajian,
  ``Analysis of the apparent lack of power in the cosmic microwave background
  anisotropy at large angular scales,''
  arXiv:astro-ph/0702723;
    E.~F.~Bunn and A.~Bourdon,
  ``Contamination cannot explain the lack of large-scale power in the cosmic
  microwave background radiation,''
  arXiv:0808.0341 [astro-ph].


\bibitem{huterer2}
  C.~J.~Copi, D.~Huterer, D.~J.~Schwarz and G.~D.~Starkman,
  ``No large-angle correlations on the non-Galactic microwave sky,''
  arXiv:0808.3767 [astro-ph].

\bibitem{vDVZ}
H.~van Dam and M.~J.~G.~Veltman,
 ``Massive And Massless Yang-Mills And Gravitational Fields,''
Nucl.\ Phys.\ B {\bf 22}, 397 (1970);
V.~I.~Zakharov,
``Linearized gravitation theory and the graviton mass,''
JETP Lett.\  {\bf 12}, 312 (1970);

\bibitem{vainshtein}
 A.~I.~Vainshtein,
 ``To the problem of nonvanishing gravitation mass,''
Phys.\ Lett.\ B {\bf 39}, 393 (1972).

\bibitem{cham1}
  J.~Khoury and A.~Weltman,
  ``Chameleon fields: Awaiting surprises for tests of gravity in space,''
  Phys.\ Rev.\ Lett.\  {\bf 93}, 171104 (2004)
  [arXiv:astro-ph/0309300];
    ``Chameleon cosmology,''
  Phys.\ Rev.\  D {\bf 69}, 044026 (2004)
  [arXiv:astro-ph/0309411].


\bibitem{cham2}
  P.~Brax, C.~van de Bruck, A.~C.~Davis, J.~Khoury and A.~Weltman,
  ``Detecting dark energy in orbit: The cosmological chameleon,''
  Phys.\ Rev.\  D {\bf 70}, 123518 (2004)
  [arXiv:astro-ph/0408415];
    ``Chameleon dark energy,''
  AIP Conf.\ Proc.\  {\bf 736}, 105 (2005)
  [arXiv:astro-ph/0410103].


\bibitem{cham3}
  S.~S.~Gubser and J.~Khoury,
  ``Scalar self-interactions loosen constraints from fifth force searches,''
  Phys.\ Rev.\  D {\bf 70}, 104001 (2004)
  [arXiv:hep-ph/0405231];
    A.~Upadhye, S.~S.~Gubser and J.~Khoury,
  ``Unveiling chameleons in tests of gravitational inverse-square law,''
  Phys.\ Rev.\  D {\bf 74}, 104024 (2006)
  [arXiv:hep-ph/0608186].

\bibitem{DGP}
  G.~R.~Dvali, G.~Gabadadze and M.~Porrati,
  ``4D gravity on a brane in 5D Minkowski space,''
  Phys.\ Lett.\ B {\bf 485}, 208 (2000)
  [arXiv:hep-th/0005016].

\bibitem{lue}
  A.~Lue, R.~Scoccimarro and G.~D.~Starkman,
  ``Probing Newton's constant on vast scales: DGP gravity, cosmic  acceleration
  and large scale structure,''
  Phys.\ Rev.\  D {\bf 69}, 124015 (2004)
  [arXiv:astro-ph/0401515].

\bibitem{mustafa}
M.~Ishak, A.~Upadhye and D.~N.~Spergel,
  ``Probing cosmic acceleration beyond the equation of state: Distinguishing
  between dark energy and modified gravity models,''
  Phys.\ Rev.\  D {\bf 74}, 043513 (2006)
  [arXiv:astro-ph/0507184].

\bibitem{koyama}
  K.~Koyama and R.~Maartens,
  ``Structure formation in the DGP cosmological model,''
  JCAP {\bf 0601}, 016 (2006)
  [arXiv:astro-ph/0511634].

\bibitem{hu}
  I.~Sawicki, Y.~S.~Song and W.~Hu,
  ``Near-horizon solution for DGP perturbations,''
  Phys.\ Rev.\  D {\bf 75}, 064002 (2007)
  [arXiv:astro-ph/0606285].

\bibitem{hu2}
  Y.~S.~Song, I.~Sawicki and W.~Hu,
  ``Large-scale tests of the DGP model,''
  Phys.\ Rev.\  D {\bf 75}, 064003 (2007)
  [arXiv:astro-ph/0606286].

\bibitem{wiley0}
   S.~Wang, L.~Hui, M.~May and Z.~Haiman,
  ``Is Modified Gravity Required by Observations? An Empirical Consistency Test
  of Dark Energy Models,''
  Phys.\ Rev.\  D {\bf 76}, 063503 (2007)
  [arXiv:0705.0165 [astro-ph]].

\bibitem{amin}
  M.~A.~Amin, R.~V.~Wagoner and R.~D.~Blandford,
  ``A sub-horizon framework for probing the relationship between the
  cosmological matter distribution and metric perturbations,''
  arXiv:0708.1793 [astro-ph].

\bibitem{song}
  Y.~S.~Song,
  ``Large Scale Structure Formation of normal branch in DGP brane world
  model,''
  Phys.\ Rev.\  D {\bf 77}, 124031 (2008)
  [arXiv:0711.2513 [astro-ph]];
    A.~Cardoso, K.~Koyama, S.~S.~Seahra and F.~P.~Silva,
  ``Cosmological perturbations in the DGP braneworld: numeric solution,''
  Phys.\ Rev.\  D {\bf 77}, 083512 (2008)
  [arXiv:0711.2563 [astro-ph]].

\bibitem{wiley}
 W.~Fang, S.~Wang, W.~Hu, Z.~Haiman, L.~Hui and M.~May,
  ``Challenges to the DGP Model from Horizon-Scale Growth and Geometry,''
  arXiv:0808.2208 [astro-ph].

\bibitem{sergei}
  S.~L.~Dubovsky and V.~A.~Rubakov,
  ``Brane-induced gravity in more than one extra dimensions: Violation of
  equivalence principle and ghost,''
  Phys.\ Rev.\  D {\bf 67}, 104014 (2003)
  [arXiv:hep-th/0212222].

\bibitem{gigashif}
  G.~Gabadadze and M.~Shifman,
  ``Softly massive gravity,''
  Phys.\ Rev.\  D {\bf 69}, 124032 (2004)
  [arXiv:hep-th/0312289].

\bibitem{oriol}
  C.~de Rham, G.~Dvali, S.~Hofmann, J.~Khoury, O.~Pujolas, M.~Redi and A.~J.~Tolley,
``Cascading Gravity: Extending the Dvali-Gabadadze-Porrati Model to Higher Dimension,"
Phys.\ Rev.\ Lett. {\bf 100}, 251603 (2008)
[arXiv:0711.2072 [hep-th]].

\bibitem{us}
  C.~de Rham, S.~Hofmann, J.~Khoury and A.~J.~Tolley,
  ``Cascading Gravity and Degravitation,''
  JCAP {\bf 0802}, 011 (2008)
  [arXiv:0712.2821 [hep-th]].

\bibitem{claudiareview}
  C.~de Rham,
  ``An Introduction to Cascading Gravity and Degravitation,''
  arXiv:0810.0269 [hep-th].

\bibitem{nemanjacharting}
  N.~Kaloper and D.~Kiley,
  ``Charting the Landscape of Modified Gravity,''
  JHEP {\bf 0705}, 045 (2007)
  [arXiv:hep-th/0703190].

\bibitem{ppf}
  W.~Hu and I.~Sawicki,
  ``A Parameterized Post-Friedmann Framework for Modified Gravity,''
  Phys.\ Rev.\  D {\bf 76}, 104043 (2007)
  [arXiv:0708.1190 [astro-ph]].


\bibitem{giaalpha}
G.~Dvali,
  ``Predictive power of strong coupling in theories with large distance
  modified gravity,''
  New J.\ Phys.\  {\bf 8}, 326 (2006)
  [arXiv:hep-th/0610013].

\bibitem{degrav}
 G.~Dvali, S.~Hofmann and J.~Khoury,
  ``Degravitation of the cosmological constant and graviton width,''
  Phys.\ Rev.\  D {\bf 76}, 084006 (2007)
  [arXiv:hep-th/0703027].

\bibitem{nonFP}
  G.~Dvali, O.~Pujolas and M.~Redi,
  ``Non Pauli-Fierz Massive Gravitons,''
  Phys.\ Rev.\ Lett.\  {\bf 101}, 171303 (2008)
  [arXiv:0806.3762 [hep-th]].

\bibitem{ddgv}
  C.~Deffayet, G.~R.~Dvali, G.~Gabadadze and A.~I.~Vainshtein,
``Nonperturbative continuity in graviton mass versus perturbative discontinuity,''
  Phys.\ Rev.\ D {\bf 65}, 044026 (2002)
  [arXiv;hep-th/0106001].

\bibitem{strong}
N.~Arkani-Hamed, H.~Georgi and M.D. Schwartz,
  ``Effective field theory for massive gravitons and gravity in theory space,''
  Annals Phys.\  {\bf 305}, 96 (2003)
  [arXiv:hep-th/0210184].

\bibitem{luty}
  M.~A.~Luty, M.~Porrati and R.~Rattazzi,
  ``Strong interactions and stability in the DGP model,''
  JHEP {\bf 0309} (2003) 029
  [arXiv:hep-th/0303116].

\bibitem{nicolis}
A.~Nicolis and R.~Rattazzi,
 ``Classical and quantum consistency of the DGP model,''
  JHEP {\bf 0406}, 059 (2004)
  [arXiv:hep-th/0404159].

\bibitem{gruz}
  A.~Gruzinov,
  ``On the graviton mass,''
  New Astron.\  {\bf 10}, 311 (2005)
  [arXiv:astro-ph/0112246].

\bibitem{por}
  M.~Porrati,
  ``Fully covariant van Dam-Veltman-Zakharov discontinuity, and absence
  thereof,''
  Phys.\ Lett.\  B {\bf 534}, 209 (2002)
  [arXiv:hep-th/0203014].

\bibitem{dgz}
 G.~Dvali, A.~Gruzinov and M.~Zaldarriaga,
``The accelerated universe and the Moon,''
  Phys.\ Rev.\ D {\bf 68}, 024012 (2003)
  [arXiv:hep-ph/0212069].

\bibitem{llr}
E. Adelberger,  Private communication;
T.W.~Murphy, Jr., E.G.~Adelberger, J.D.~Strasburg and C.W.~Stubbs,
`APOLLO: Multiplexed Lunar Laser Ranging', at
{\tt http://physics.ucsd.edu/$\sim$tmurphy/
apollo/doc/multiplex.pdf}.

\bibitem{battat}
  J.~B.~R.~Battat, C.~W.~Stubbs and J.~F.~Chandler,
  ``Solar system constraints on the Dvali-Gabadadze-Porrati braneworld theory
  of gravity,''
  Phys.\ Rev.\  D {\bf 78}, 022003 (2008)
  [arXiv:0805.4466 [gr-qc]].

\bibitem{cedric}
C.~Deffayet,
``Cosmology on a brane in Minkowski bulk,''
Phys.\ Lett.\  {\bf B 502}, 199 (2001)
[arXiv:hep-th/0010186].

\bibitem{koyamaghost}
  K.~Koyama,
  ``Are there ghosts in the self-accelerating brane universe?,''
  Phys.\ Rev.\  D {\bf 72}, 123511 (2005)
  [arXiv:hep-th/0503191];
    D.~Gorbunov, K.~Koyama and S.~Sibiryakov,
  ``More on ghosts in DGP model,''
  Phys.\ Rev.\  D {\bf 73}, 044016 (2006)
  [arXiv:hep-th/0512097].

\bibitem{ruth}
 C.~Charmousis, R.~Gregory, N.~Kaloper and A.~Padilla,
  ``DGP specteroscopy,''
  JHEP {\bf 0610}, 066 (2006)
  [arXiv:hep-th/0604086].

\bibitem{dw}
  G.~Dvali, G.~Gabadadze, O.~Pujolas and R.~Rahman,
  ``Domain walls as probes of gravity,''
  Phys.\ Rev.\  D {\bf 75}, 124013 (2007)
  [arXiv:hep-th/0612016].

\bibitem{rob}
  R.~Gregory, N.~Kaloper, R.~C.~Myers and A.~Padilla,
  ``A New Perspective on DGP Gravity,''
  JHEP {\bf 0710}, 069 (2007)
  [arXiv:0707.2666 [hep-th]].

\bibitem{turner}
  G.~Dvali and M.~S.~Turner,
  ``Dark energy as a modification of the Friedmann equation,''
  arXiv:astro-ph/0301510.

\bibitem{claudiaandrew}
C.~de Rham and A.~J.~Tolley, private communication.

\bibitem{w<-1}
 A.~Lue and G.~D.~Starkman,
  ``How a brane cosmological constant can trick us into thinking that $w <
  -1$,''
  Phys.\ Rev.\  D {\bf 70}, 101501 (2004)
  [arXiv:astro-ph/0408246];
    V.~Sahni and Y.~Shtanov,
  ``Braneworld models of dark energy,''
  JCAP {\bf 0311}, 014 (2003)
  [arXiv:astro-ph/0202346];
   L.~P.~Chimento, R.~Lazkoz, R.~Maartens and I.~Quiros,
  ``Crossing the phantom divide without phantom matter,''
  JCAP {\bf 0609}, 004 (2006)
  [arXiv:astro-ph/0605450].

\bibitem{DMDE1}
    S.~M.~Carroll, A.~De Felice and M.~Trodden,
  ``Can we be tricked into thinking that w is less than -1?,''
  Phys.\ Rev.\  D {\bf 71}, 023525 (2005)
  [arXiv:astro-ph/0408081];

\bibitem{DMDE2}
  S.~Das, P.~S.~Corasaniti and J.~Khoury,
  ``Super-acceleration as signature of dark sector interaction,''
  Phys.\ Rev.\  D {\bf 73}, 083509 (2006)
  [arXiv:astro-ph/0510628].

\bibitem{staro}
  B.~Boisseau, G.~Esposito-Farese, D.~Polarski and A.~A.~Starobinsky,
  ``Reconstruction of a scalar-tensor theory of gravity in an accelerating
  universe,''
  Phys.\ Rev.\ Lett.\  {\bf 85}, 2236 (2000)
  [arXiv:gr-qc/0001066].


\bibitem{bhuvnesh}
  B.~Jain and P.~Zhang,
  ``Observational Tests of Modified Gravity,''
  arXiv:0709.2375 [astro-ph].

\bibitem{ed2}
  E.~Bertschinger and P.~Zukin,
  ``Distinguishing Modified Gravity from Dark Energy,''
  Phys.\ Rev.\  D {\bf 78}, 024015 (2008)
  [arXiv:0801.2431 [astro-ph]].

\bibitem{caldwell}
 S.~F.~Daniel, R.~R.~Caldwell, A.~Cooray and A.~Melchiorri,
  ``Large Scale Structure as a Probe of Gravitational Slip,''
  Phys.\ Rev.\  D {\bf 77}, 103513 (2008)
  [arXiv:0802.1068 [astro-ph]].

\bibitem{koyamaparam}
  K.~Koyama,
  ``Structure formation in modified gravity models alternative to dark
  energy,''
  JCAP {\bf 0603}, 017 (2006)
  [arXiv:astro-ph/0601220].

\bibitem{separate}
  D.~H.~Lyth, K.~A.~Malik and M.~Sasaki,
  ``A general proof of the conservation of the curvature perturbation,''
  JCAP {\bf 0505}, 004 (2005)
  [arXiv:astro-ph/0411220].

\bibitem{ed}
 E.~Bertschinger,
  ``On the Growth of Perturbations as a Test of Dark Energy,''
  Astrophys.\ J.\  {\bf 648}, 797 (2006)
  [arXiv:astro-ph/0604485].

\bibitem{Sachs:1967er}
  R.~K.~Sachs and A.~M.~Wolfe,
  ``Perturbations of a cosmological model and angular variations of the
  microwave background,''
  Astrophys.\ J.\  {\bf 147}, 73 (1967).

\bibitem{camb}
  A.~Lewis, A.~Challinor and A.~Lasenby,
  ``Efficient Computation of CMB anisotropies in closed FRW models,''
  Astrophys.\ J.\  {\bf 538}, 473 (2000)
  [arXiv:astro-ph/9911177].

\bibitem{cambppf}
W.~Fang, {\tt http://camb.info/ppf/}.


\bibitem{Komatsu}
  E.~Komatsu {\it et al.}  [WMAP Collaboration],
  ``Five-Year Wilkinson Microwave Anisotropy Probe (WMAP)
  Observations: Cosmological Interpretation,''
  arXiv:0803.0547 [astro-ph].

\bibitem{Spergel:2003cb}
  D.~N.~Spergel {\it et al.}  [WMAP Collaboration],
  ``First Year Wilkinson Microwave Anisotropy Probe (WMAP) Observations:
  Determination of Cosmological Parameters,''
  Astrophys.\ J.\ Suppl.\  {\bf 148}, 175 (2003)
  [arXiv:astro-ph/0302209].

\bibitem{Nolta:2008ih}
  M.~R.~Nolta {\it et al.}  [WMAP Collaboration],
  ``Five-Year Wilkinson Microwave Anisotropy Probe (WMAP) Observations: Angular
  Power Spectra,''
  arXiv:0803.0593 [astro-ph].

  \bibitem{Planck}
    [Planck Collaboration],
  ``Planck: The scientific programme,''
  arXiv:astro-ph/0604069.

\bibitem{Hui:2007zh}
  L.~Hui and K.~P.~Parfrey,
  ``The Evolution of Bias - Generalized,''
  Phys.\ Rev.\  D {\bf 77}, 043527 (2008)
  [arXiv:0712.1162 [astro-ph]].

  \bibitem{Fu:2007qq}
  L.~Fu {\it et al.},
  ``Very weak lensing in the CFHTLS Wide: Cosmology from cosmic shear in the
  linear regime,''
  arXiv:0712.0884 [astro-ph].

\bibitem{LSST}
{\tt http://www.lsst.org/lsst}

\bibitem{panstarrs}
{\tt  http://pan-starrs.ifa.hawaii.edu/}

\bibitem{JDEM}
{\tt http://jdem.gsfc.nasa.gov/}

 \bibitem{Crittenden:1995ak}
  R.~G.~Crittenden and N.~Turok,
  ``Looking for $\Lambda$ with the Rees-Sciama Effect,''
  Phys.\ Rev.\ Lett.\  {\bf 76}, 575 (1996)
  [arXiv:astro-ph/9510072].

\bibitem{Neyrinck}
  B.~R.~Granett, M.~C.~Neyrinck and I.~Szapudi,
  arXiv:0812.1025 [astro-ph].

   \bibitem{Afshordi:2004kz}
  N.~Afshordi,
  ``Integrated Sachs-Wolfe effect in Cross-Correlation: The Observer's
  Manual,''
  Phys.\ Rev.\  D {\bf 70}, 083536 (2004)
  [arXiv:astro-ph/0401166].

\bibitem{Viel:2005ha}
  M.~Viel and M.~G.~Haehnelt,
   ``Cosmological and astrophysical parameters from the SDSS flux power spectrum
  and hydrodynamical simulations of the Lyman-alpha forest,''
  Mon.\ Not.\ Roy.\ Astron.\ Soc.\  {\bf 365}, 231 (2006)
  [arXiv:astro-ph/0508177].

  \bibitem{Dawson:2006qd}
  K.~S.~Dawson, W.~L.~Holzapfel, J.~E.~Carlstrom, M.~Joy and S.~J.~LaRoque,
  ``Final Results from the BIMA CMB Anistropy Survey and Search for Signature
  of the SZ effect,''
  Astrophys.\ J.\  {\bf 647}, 13 (2006)
  [arXiv:astro-ph/0602413].

  \bibitem{Komatsu:2002wc}
  E.~Komatsu and U.~Seljak,
  ``The Sunyaev-Zel'dovich angular power spectrum as a probe of cosmological
  parameters,''
  Mon.\ Not.\ Roy.\ Astron.\ Soc.\  {\bf 336}, 1256 (2002)
  [arXiv:astro-ph/0205468].

  \bibitem{Afshordi:2006pc}
  N.~Afshordi, Y.~T.~Lin, D.~Nagai and A.~J.~R.~Sanderson,
  ``Missing Thermal Energy of the Intracluster Medium,''
  Mon.\ Not.\ Roy.\ Astron.\ Soc.\  {\bf 378}, 293 (2007)
  [arXiv:astro-ph/0612700].
  
  \bibitem{vikhlinin}
  A.~Vikhlinin {\it et al.},
  ``Chandra Cluster Cosmology Project III: Cosmological Parameter
  Constraints,''
  arXiv:0812.2720 [astro-ph].

\bibitem{vikhlinin_analysis}
  A.~Vikhlinin {\it et al.},
  ``Chandra Cluster Cosmology Project II: Samples and X-ray Data Reduction,''
  arXiv:0805.2207 [astro-ph].

\bibitem{mark}
J.~Khoury and M.~Wyman, to appear.




\end{thebibliography}
\end{document}